\documentclass[twocolumn,times]{aastex631}
\usepackage{amsmath}
\begin{document}

\newcommand{\red}[1]{\textcolor{red}{#1}}
\newcommand{\blue}[1]{\textcolor{blue}{#1}}
\newcommand{\cvec}[1]{\left[#1\right]_\mathrm{car}}
\newcommand{\svec}[1]{\left[#1\right]_\mathrm{sph}}
\newcommand{\dd}{{\rm d}}
\newcommand{\nsp}{n_\mathrm{sp}}
\newcommand{\A}{{\cal A}}
\newcommand{\eps}{\varepsilon}
\renewcommand{\phi}{\varphi}
\newcommand{\tet}{\vartheta}
\newcommand{\btet}{\Theta}
\newcommand{\waz}{w_a(z)}
\newcommand{\p}{^\prime}
\newcommand{\com}{^\mathrm{C}}
\newcommand{\inc}{^\mathrm{I}}
\newcommand{\sca}{^\mathrm{sc}}
\newcommand{\za}{{z_\mathrm{a}}}
\newcommand{\betaren}{{\beta_\mathrm{ren}}}
\newcommand{\colvec}[2]{\left(\begin{array}{c} #1 \\ #2\end{array} \right)}

\newcommand{\PD}[2]{\frac{\partial #1}{\partial #2}}
\title{\large An exact analytical solution for the weakly magnetized flow around an axially symmetric paraboloid, with application to magnetosphere models}
\author[0000-0001-6122-9376]{Jens Kleimann}
\affiliation{Theoretische Physik IV, Ruhr-Universit\"at Bochum, 44780 Bochum, Germany}
\affiliation{Ruhr Astroparticle and Plasma Physics Center (RAPP Center), Germany} 
\email{e-mail: jk@tp4.rub.de}
\author[0000-0001-5490-8581]{Christian R\"oken}
\affiliation{Department of Geometry and Topology, Faculty of Science, University of Granada, 18071 Granada, Spain}
\affiliation{Lichtenberg Group for History and Philosophy of Physics, Institut f\"ur Philosophie, Universit\"at Bonn, 53115 Bonn, Germany}

\begin{abstract}
\noindent Rotationally symmetric bodies with longitudinal cross sections of parabolic shape are frequently used to model astrophysical objects, such as magnetospheres and other blunt objects, immersed in interplanetary or interstellar gas or plasma flows.
We discuss a simple formula for the potential flow of an incompressible fluid around an elliptic paraboloid whose axis of symmetry coincides with the direction of incoming flow.
Prescribing this flow, we derive an exact analytical solution to the induction equation of ideal magnetohydrodynamics for the case of an initially homogeneous magnetic field of arbitrary orientation being passively advected in this flow. Our solution procedure employs Euler potentials and Cauchy's integral formalism based on the flow's stream function and isochrones.
Furthermore, we use a particular renormalization procedure that allows us to generate more general analytical expressions modeling the deformations experienced by arbitrary scalar or vector-valued fields embedded in the flow as they are advected first toward and then past the parabolic obstacle. 
Finally, both the velocity field and the magnetic field embedded therein are generalized from incompressible to mildly compressible flow, where the associated density distribution is found from Bernoulli's principle.
\end{abstract}

\vspace*{2mm}

\section{Introduction}
\label{sec:intro}

\noindent The study of fluid flow around solid obstacles is a classical topic of hydrodynamics, with countless applications in both engineering and physics, and especially astrophysics. Therein, the case of the fluid being magnetized is of particular interest for several classes of objects, ranging from the heliosphere \citep[e.g.,][]{Roeken_EA:2015, Kleimann_EA:2022}, planets \citep[e.g.,][]{Petrinec_Russell:1997, Kotova_EA:2020} and their magnetospheres \citep[e.g.,][]{Spreiter_Stahara:1995, Richardson:2002}, comets and planetary satellites \citep[e.g.,][and references therein]{Combi_Gombosi:2002}, up to clusters of galaxies \citep[e.g.,][]{Dursi_Pfrommer:2008}.

While quantitative applications are often faced with complications such as turbulence, reactive flows, or high Mach numbers that typically can only be investigated using numerical simulations, a limited set of analytical models has been obtained by restriction to stationary flows that exhibit some form of spatial symmetry. For instance, \citet{Nabert_EA:2013} employ certain axial and point-like symmetries of the magnetohydrodynamic (MHD) quantities and a stationary flow to derive the plasma properties of the Earth's magnetosheath, the region enclosed between the magnetopause and the bow shock. In a similar vein, the present paper considers stationary flows $\vec{u}$ which satisfy the constraints of being both irrotational
\begin{equation}
  \label{eq:irrotational}
  \nabla \times \vec{u} = \vec{0}
\end{equation}
and solenoidal
\begin{equation}
  \label{eq:solenoidal}
  \nabla \cdot \vec{u} = 0 \, .
\end{equation}
The popularity of these two constraints is mainly due to the fact that they allow $\vec{u}$ to be expressed through a scalar potential $\Phi$ via $\vec{u} = -\nabla \Phi$, with $\Phi$ satisfying the Laplace equation
\begin{equation}
  \label{eq:laplace}
  \nabla^2 \Phi = \nabla \cdot (\nabla \Phi) = \nabla \cdot (-\vec{u}) = 0 \, .
\end{equation}
This implies in particular that the flow field is fully determined by the shape of the obstacle and the flow conditions prevailing at the upstream boundary.
In the case of planar symmetry, in which the shapes and flows are invariant along a Cartesian direction, classical examples of obstacles for which such potential flows have been explicitly derived include infinite planes, wedges, and the circular cylinder. By introducing a complex velocity potential, it is possible to generate additional solutions from the existing ones through the application of conformal mappings such as the Joukowski transform \citep[e.g.,][p.\ 435]{Batchelor:2000}.

This paper, however, is concerned with settings where both the solid object and the flow around it feature an axial symmetry with respect to the inflow direction, rather than a translational one.
For these, the method of complex potentials and conformal mappings is not available, and thus the set of known examples is correspondingly smaller.
The set of known nontrivial solutions for magnetized generalizations of such flows is smaller still, having until now essentially been restricted to the sphere \citep[e.g.,][and references therein]{Dursi_Pfrommer:2008, Isenberg_EA:2015} and the Rankine-type heliosphere \citep{Roeken_EA:2015, Isenberg_EA:2015}. In the former case of a sphere, the exact magnetic field solution is known only up to an integral which so far could not be expressed in terms of analytical functions, whereas the latter Rankine-type case involves special functions, namely, incomplete elliptic integrals. In contrast to these, the MHD solutions to be presented here are predominately of rather compact form and expressible through basic functions without recourse to unevaluated integrals, making them a viable alternative to numerical computations at least in simple applications.

Moreover, besides the obvious puristic and educational motivations, this geometrical setting is also of significant astrophysical interest. 
Starting with the work of \citet{Kobel_Fluckiger:1994} and its subsequent refinements \citep[e.g.,][and references therein]{Romashets_Vandas:2019}, the upwind sides of planetary magnetopauses continue to be frequently modeled as rotationally symmetric paraboloids. 
More generally, the situation of a flow impacting on a solid, convex object is obviously a very generic one. It is thus also potentially applicable to the interaction of other objects like astrospheres and their galactic analogs with their respective magnetized environments. Particularly 
the latter has received relatively little attention since the works of \citet{Ikeuchi_Tomisaka:1981} but is becoming increasingly relevant in the light of more recent observations and associated modeling of cosmic rays accelerated at a Galactic termination shock \citep{Merten_EA:2018}, as well as works on galaxy-scale interactions such as ram-pressure stripping \citep[e.g.,][]{Ramos-Martinez_EA:2018, Mueller_EA:2021}.

It is clear that the numerous simplifying assumptions inherent to analytical approaches of the present kind, such as the stationarity and homogeneity of the incident flow, the perfectly regular shape of the geometric surface, and the absence of turbulent small-scale motion, will typically render them less suitable for an accurate reproduction of real-world astrophysical scenarios, which are therefore better left to numerical investigations.
Then again, it is exactly this simplifying approach that allows for the model to be applied to such a relatively wide range of different (astro)physical scenarios in the first place.
And while the present approach cannot include discontinuous surfaces like bow shocks because it computes its velocity components from a smooth potential, it is certainly relevant at least in situations where a bow shock can either be neglected or is absent altogether.

The paper is structured as follows. After this introduction, the analytical expressions and properties of the potential flow field around an axially symmetric paraboloid are derived and discussed in Section~\ref{sec:flow}. Thereafter, Section~\ref{sec:magnetic} presents the analytical derivation and discussion of a magnetic field $\vec{B}$ being passively advected in this flow by solving the stationary induction equation of ideal MHD,
\begin{equation}
  \label{eq:induct}
  \nabla \times ( \vec{u} \times \vec{B} ) = \vec{0} \, .
\end{equation}
This includes the extension to paraboloids being linearly scaled to a custom aspect ratio.
Section~\ref{sec:apply} presents further generalizations, most notably to the advection of more general scalar and vectorial fields.
Finally, Section~\ref{sec:compressible} generalizes the ambient paraboloid flow field to subsonic compressible flow, including the associated distribution of mass density, and derives the correspondingly changed magnetic field, after which Section~\ref{sec:summary} concludes the paper with a summary. \\

\section{The inviscid flow field}
\label{sec:flow}

\noindent In order to motivate the course of the present analysis, we start by considering the example of inviscid flow around a sphere, whose well-known flow potential and stream function \citep[e.g.,][p.~452]{Batchelor:2000} in terms of cylindrical coordinates $(\rho, \phi, z) \in \mathbb{R}_{\ge 0} \times [0, 2 \pi) \times \mathbb{R}$ read, respectively,
\begin{align}
  \label{eq:phi_sphere}
  \Phi_0 &= z \left( 1+\frac{1}{2 \, r^3} \right) \\[0.20cm]
  \label{eq:psi_sphere}
  \Psi_0 &= -\frac{\rho^2}{2} \left( 1-\frac{1}{r^3} \right) ,
\end{align}
and result in the flow field
\begin{align}
    \label{eq:u_sphere}
   \nonumber \vec{u}_0 &= - \nabla \Phi_0
  = \nabla \times \left[(\Psi_0/\rho) \, \vec{e}_\phi \right] \\[0.20cm]
  &= \frac{3 \, \rho \, z}{2 \, r^5} \, \vec{e}_\rho
  + \left(\frac{z^2 - \rho^2/2}{r^5} - 1\right) \vec{e}_z \, ,
\end{align}
where we use the shorthand $r=\sqrt{\rho^2+z^2}$ wherever it allows for more compact notation of formulas. Throughout the paper, quantities of lengths are normalized to a characteristic length scale (such as the sphere's radius in the above case), while all other quantities, most notably the magnitudes of the velocity and the magnetic field, are normalized to their respective values at the upstream boundary.

Since streamlines are given by lines of constant $\Psi_0$, we may interpret Equation~(\ref{eq:psi_sphere}) as a special case of the simpler stream function $\Psi_\mathrm{bg} := -\rho^2/2$ of the undisturbed background flow \mbox{$\vec{u}_\mathrm{bg} = -\vec{e}_z$} being modulated by a factor $1-1/f(\rho, z)$ that characterizes the shape of the obstacle in such a way that $f(\rho, z) = 1$ represents its contour, in this case the sphere with $r = 1$.
The particular streamline $\Psi_0=0$ will then include not only the entire $z$-axis but split at two stagnation points (or one in cases where the obstacle is of infinite extent in the downstream direction), from which the emanating streamlines precisely trace the obstacle's surface.

This observation invites the question whether the flow around other shapes with axial symmetry might be obtained in an analogous manner. For instance, one might attempt to generalize the sphere flow with its circular contour \mbox{$f(\rho, z)=r^3$} to the more general contour 
\begin{equation}
   \label{eq:ansatz}
   f(\rho, z) = g_\eps(r +\eps \, z)
\end{equation}
with $g_0(r) = r^3$ and the argument 
encoding the parametric equation
\begin{equation}
  \label{eq:conic}
   r +\eps \, z = 1 \quad \Leftrightarrow \quad r = \frac{1}{1+\eps \cos\tet}
\end{equation}
of a conic section with its focal point at the origin, where $z = r \cos \tet$ and $\tet \in (0, \pi)$, covering ellipses ($0 < \eps < 1$) and hyperbolas ($\eps > 1$) of eccentricity $\eps$, as well as the special cases of circle ($\eps = 0$) and parabola ($\eps = 1$). However, while the resulting flows around such ellipsoidal or hyperboloidal bodies of revolution are indeed solenoidal by construction (and might therefore be of actual interest for specialized applications), direct calculations show that no $g_\eps(\cdot)$ that additionally satisfies Equation~(\ref{eq:irrotational}) exists for such bodies (see Appendix~\ref{sec:appA}). On the other hand, similar considerations can be used to generalize the sphere's flow potential (\ref{eq:phi_sphere}) to nonzero eccentricities, resulting in flow fields satisfying Equation~(\ref{eq:irrotational}) but not (\ref{eq:solenoidal}).
We note that the correct components for flow around ellipsoids satisfying both constraints
involve the solution of the Laplace equation $\nabla^2 \Phi_\eps=0$ in ellipsoidal coordinates \citep[e.g.,][p.~475]{Milne-Thomson:1968}, and the rather complicated form of the associated expressions renders the general case $0<\eps<1$ less attractive in the present context.

\begin{figure*}
  \centering
  \includegraphics[width=\textwidth]{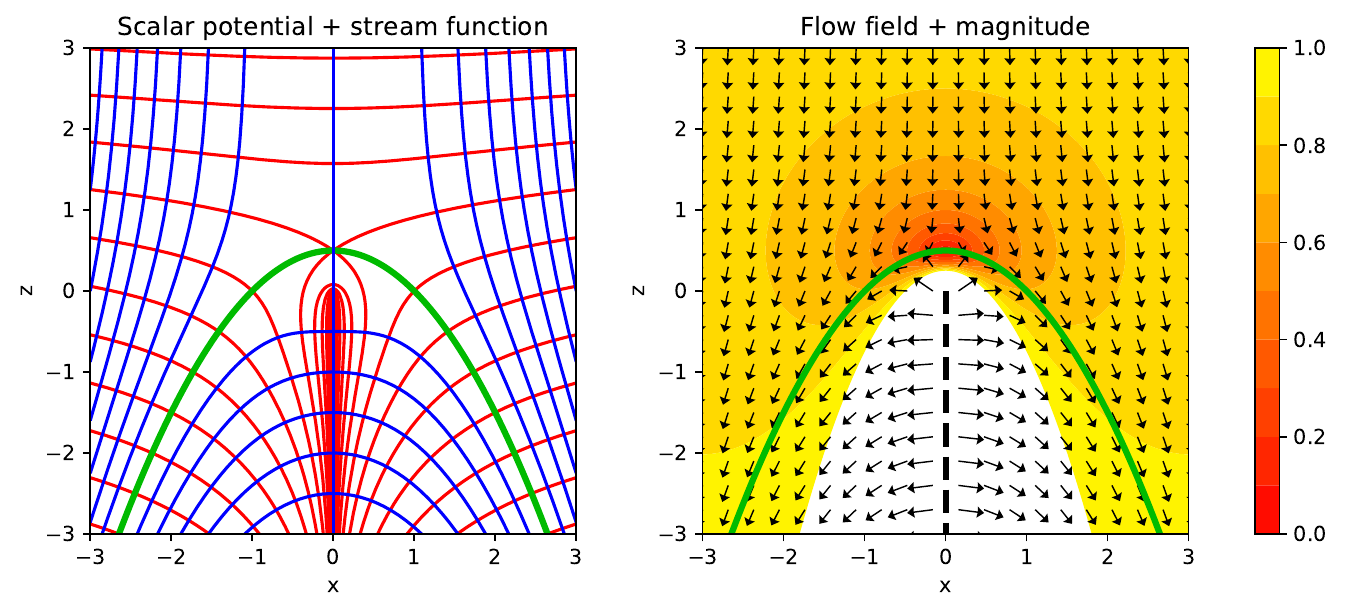}
  \caption{ \label{fig:flowfield}
    Left: streamlines (lines of constant $\Psi$, blue) and equipotentials
    (lines of constant $\Phi$, red) at 0.5 spacing, with the parabolic obstacle (green) given by its parametric form (\ref{eq:para}).
    Right: the resulting vector field represented by arrows and contours of absolute speed $\|\vec{u}\|$. The dashed black line marks the flow-generating, semi-infinite line source.
    Arrow lengths have been scaled $\propto \|\vec{u}\|^{0.3}$ in order to accommodate high speeds near the source, and contours above unity have been suppressed since these occur exclusively inside the obstacle. \\
  }
\end{figure*}

In the following, we will therefore focus on the parabolic case $\eps=1$ with $g_1(r + z) = r + z$ being the identity mapping. This leads to the stream function
\begin{equation}
  \label{eq:psi_para}
  \Psi = -\frac{\rho^2}{2} \left( 1-\frac{1}{r + z} \right) = - \frac{1}{2} \, (r - z) \, (r + z - 1) \, ,
\end{equation}
where the parabolic obstacle itself is given by
\begin{equation}
  \label{eq:para}
  r + z = 1 \quad \Leftrightarrow \quad 2 \, z = 1-\rho^2 \ ,
\end{equation}
and the lower index ``1,'' which was previously used to hint at $\eps=1$, will be suppressed from now on.
Equation~(\ref{eq:psi_para}) reproduces the stream function formula
\begin{equation}
  \psi = 2\, c^2 U \left(\xi^2 - \xi_0^2 \right) \eta^2 
\end{equation}
from \citet[][p.~479]{Milne-Thomson:1968}, expressed in elliptical coordinates $(\xi,\eta)$ satisfying
$(\rho,z)= \left( 2\, c \, \xi \, \eta, [\xi^2-\eta^2] \, c \right)$ when choosing $c=U=1$ and $\xi_0=1/\sqrt{2}$, up to an overall minus sign (i.e., $\psi = -\Psi$) that originates from an alternative definition of the stream function used therein. Applying the relation
\begin{equation}
    \label{eq:u_from_Psi}
    \vec{u} = \nabla \times \left[ (\Psi/\rho) \, \vec{e}_\phi \right]
\end{equation}
to the stream function (\ref{eq:psi_para}), we obtain flow components
\begin{align}
  \label{eq:u_rho_from_Psi} u_{\rho} &= -\frac{1}{\rho} \PD{\Psi}{z} = \frac{1}{2 \, r} \, \sqrt{\frac{r - z}{r + z}} \\[0.20cm]
  \label{eq:u_z_from_Psi} u_z &= +\frac{1}{\rho} \PD{\Psi}{\rho} = \frac{1}{2 \, r} -1  
\end{align}
which, by way of direct calculation, can indeed be shown to be irrotational (see again Appendix~\ref{sec:appA}). Therefore, a flow potential for this field exists and is easily found to be of the simple form
\begin{equation}
  \label{eq:phi_para}
  \Phi = z - \frac{\ln (r + z)}{2}  .
\end{equation}
Figure~\ref{fig:flowfield} visualizes the structure of the flow field~(\ref{eq:u_rho_from_Psi})--(\ref{eq:u_z_from_Psi}) and the associated stream function (\ref{eq:psi_para}) and potential (\ref{eq:phi_para}). As can be seen from both the plots and Equation~(\ref{eq:psi_para}), the parabolic surface comes about as the separatrix between the incident background flow and that of a line source extending along the entire negative $z$-axis, in contrast to the sphere case, where the flow is created by a point dipole at the origin. 
As expected, the stagnation point is located at $z=1/2$. Along the stagnation line ($\rho=0, z>1/2$), the flow decelerates rather gently according to
\mbox{$u_z|_{\rho=0} = (2 \, z)^{-1} -1$},
unlike the flow ahead of a sphere, for which Equation~(\ref{eq:u_sphere}) yields a somewhat sharper profile, namely \mbox{$u_{0,z}|_{\rho=0} = z^{-3} -1$}. This difference is likely caused by the paraboloid's infinite lateral extent, which requires a much larger body of fluid to be displaced.

The flow field~(\ref{eq:u_rho_from_Psi})--(\ref{eq:u_z_from_Psi}) is of course markedly simpler than the corresponding viscous flow that has been investigated analytically by \citet{Miller:1969, Miller:1971} and numerically by, e.g., \citet{Davis_Werle:1972} because it avoids the intricate modeling of boundary layers that is required when solving the Navier--Stokes equation for the geometry in question. \\


\section{The frozen-in magnetic field}
\label{sec:magnetic}

\subsection{The longitudinal and transversal components}

\noindent Next, we turn to the computation of the components of a magnetic field that is embedded into the flow (\ref{eq:u_rho_from_Psi})--(\ref{eq:u_z_from_Psi}) and passively advected in it. As often in such cases, we require the magnetic field at upstream infinity ($z \rightarrow +\infty$) to be of constant (though arbitrary) orientation and magnitude. The boundary field is best specified through its constant Cartesian components $(B_{x 0}, B_{y 0}, B_{z 0})$, yielding the cylindrical representation
\begin{align}
  \label{eq:Bo_rho} B_{\rho 0} :=& \lim_{z \rightarrow \infty}B_{\rho} = +B_{x 0} \cos\phi + B_{y 0} \sin\phi \\[0.20cm]
  \label{eq:Bo_phi} B_{\phi 0} :=& \lim_{z \rightarrow \infty}B_{\phi} = -B_{x 0} \sin\phi + B_{y 0} \cos\phi \\[0.20cm]
  \label{eq:Bo_z} B_{z 0} =& \lim_{z \rightarrow \infty} B_z \, .
\end{align}

To derive the full frozen-in magnetic field, we closely follow the procedure used by \citet{Isenberg_EA:2015} for the Rankine half-body flow by employing the decomposition
\begin{equation}
  \label{eq:split_B}
  \vec{B} = \vec{B}_\mathrm{lo} + \vec{B}_\mathrm{tr}
\end{equation}
of the magnetic field into a longitudinal (flow-parallel) component $\vec{B}_\mathrm{lo}$ and a transversal component $\vec{B}_\mathrm{tr}$, which at the upstream boundary (but generally not at other locations) is perpendicular to the flow.
When introducing the azimuthal angle $\phi_0$ of the boundary field, i.e., the unique angle satisfying
\begin{equation}
  \label{eq:Bxy_0}
  \frac{B_{x 0}}{\cos\phi_0} = \frac{B_{y 0}}{\sin\phi_0} =
  \sqrt{ (B_{x 0})^2 + (B_{y 0})^2 } = B_\mathrm{tr,0} \, ,
\end{equation}
the two horizontal components (\ref{eq:Bo_rho}) and (\ref{eq:Bo_phi}) may alternatively be written as
\begin{align}
  B_{\rho 0} &= + B_{\mathrm{tr},0} \cos(\phi-\phi_0) \\[0.20cm]
  B_{\phi 0} &= - B_{\mathrm{tr},0} \sin(\phi-\phi_0) \, .
\end{align}
Now, being anti-parallel to $\vec{u}$, which at upstream infinity is just
$\lim_{z \rightarrow \infty}\vec{u} = -\vec{e}_z$, the longitudinal component satisfying the boundary condition~(\ref{eq:Bo_z}) is trivially found to be
\begin{align}
  \label{eq:b_long}
  \vec{B}_\mathrm{lo} &= B_{z 0} \, [-\vec{u}] \\[0.20cm]
  &= B_{z 0} \left[ - \frac{1}{2 \, r} \, \sqrt{\frac{r - z}{r + z}} \, \vec{e}_\rho + \left( 1-\frac{1}{2 \, r} \right) \vec{e}_z \right] . \nonumber
\end{align}

And just like any solenoidal magnetic field, the transverse component can be written as
\begin{equation}
  \label{eq:euler}
  \vec{B}_\mathrm{tr} = \nabla \alpha \times \nabla \beta
\end{equation}
using Euler potentials $\alpha$ and $\beta$ \citep[e.g.,][]{Stern:1966}, which are scalar functions of position.
This ansatz implies in particular that field lines may be identified as the lines at which surfaces of constant $\alpha$ intersect those of constant $\beta$. Furthermore, it guarantees that the divergence constraint $\nabla \cdot \vec{B}_\mathrm{tr}=0$ is always satisfied.
Here, we choose the first Euler potential $\alpha$ as
\begin{align}
  \label{eq:euler_alpha} \nonumber
  \alpha &= B_\mathrm{tr,0} \, \sqrt{-2 \, \Psi} \, \sin (\phi - \phi_0) \\[0.20cm]
  &= B_\mathrm{tr,0} \, \sqrt{(r - z) \, (r + z - 1)} \, \sin (\phi - \phi_0)
\end{align}
in complete analogy to Equation~(7) in \citet{Isenberg_EA:2015}.
This ensures that (i) a streamline cannot leave its surface of constant $\alpha$ and that (ii) said surfaces thus form an infinite bundle of nonintersecting sheets that are parallel to the boundary field and to each other at infinity, but then drape around the parabolic obstacle like a vertical curtain.

\subsection{Isochrones} \label{Sec-3-2}

\noindent The second Euler potential $\beta$ is chosen as the negative of the travel time $t$ that a fluid particle spent moving from upstream infinity along a trajectory $t \mapsto \big(\rho(t), z(t)\big)$ to its current position, and lines (or, in three dimensions, surfaces) of constant $\beta$ will therefore be referred to as isochrones.
The minus sign, which is absent from the formulas used by \citet{Isenberg_EA:2015}, has to be added to either $\alpha$ or $\beta$ to ensure consistency with the desired boundary conditions (\ref{eq:Bo_rho})--(\ref{eq:Bo_z}), and we chose $\beta$ for reasons that will become clear in Section~\ref{sec:general_b}.

From the definitions of the velocity components $u_\rho$ and $u_z$, which then read
\begin{equation}
  \left( u_\rho, u_z \right) =
  \left( -\frac{\dd \rho(\beta)}{\dd \beta}, -\frac{\dd z(\beta)}{\dd \beta} \right) ,
\end{equation}
we obtain the respective formulas for $\beta$ as
\begin{equation}
  \label{eq:def_beta}
  -\int_a^\rho \frac{\dd \rho\p}{u_\rho(\rho\p)|_{\Psi}} = \beta =
  -\int_{\infty}^z \frac{\dd z\p }{u_z(z\p)|_{\Psi}} .
\end{equation}
The symbol $\Psi$ in the integrand indicates that integration has to be performed along a streamline, i.e., at constant $\Psi$, and $a$ is the value attained by the $\rho$ coordinate of the streamline in question at $z \rightarrow \infty$, that is, the value satisfying
\begin{equation}
  \label{eq:rel_a-Psi} 
  \Psi(\rho,z) = \Psi(a, \infty) = -a^2/2 \ .
\end{equation}
This results in the explicit expression
\begin{equation}
  \label{eq:def_a}
  a(\rho,z) = \sqrt{-2 \, \Psi(\rho,z)}
  = \sqrt{(r-z) \, (r+z-1)}
\end{equation}
and allows us to use $a$ rather than $\Psi$ to label streamlines.

We point out that Equation~(\ref{eq:def_beta}) suffers from the conceptual problem that it will necessarily lead to infinite travel times for any $(\rho,z)$, consistent with an infinitely long distance being traversed at finite speed. Actual computations, such as that of the resulting magnetic field components via Equation~(\ref{eq:euler}), therefore require the integral to start at a finite height $z_0$ (and a corresponding axial distance $\rho_0$) to keep the travel time finite. Only afterward may one perform the limit $(\rho_0, z_0) \rightarrow (a, \infty)$ along the streamline in question. (See Section~\ref{sec:isochrones} for a more sophisticated remedy.)

While both expressions in Equation~(\ref{eq:def_beta}) are fully equivalent from a mathematical point of view, tentative computations indicate that the first expression should be preferred for reasons of simplicity. We may thus solve Equation~(\ref{eq:psi_para}) for $z$ to obtain
\begin{equation}
  \label{eq:z-rhopsi}
  2 \, z = \frac{\rho^2}{\rho^2 + 2 \, \Psi} - \rho^2 - 2 \, \Psi 
  = \frac{\rho^2}{\rho^2 - a^2} - \rho^2 + a^2
\end{equation}
and, after a series of elementary simplifications, are lead to the indefinite integral
\begin{align} \nonumber
  \label{eq:integral}
  T(\rho, a) &:= \int \frac{\dd \rho}{u_\rho(\rho, a)}
  = \int \left(1 +\frac{\rho^2}{(\rho^2 - a^2)^2} \right) \rho \, \dd \rho \\[0.20cm]
  &= \frac{\rho^2 - a^2}{2} + \frac{\ln \left(\rho^2 - a^2\right)}{2} 
  - \frac{a^2}{2 \, (\rho^2 - a^2) } \, .
\end{align}
It is both noteworthy and fortunate that this exact integral can be solved easily and expressed without recourse to special functions such as incomplete elliptic integrals, as is the case for the Rankine heliosphere \citep{Roeken_EA:2015}.
The travel time of a fluid element moving from $(\rho_0, z_0)$ to $(\rho,z)$ along a streamline $a=a(\rho_0,z_0)=a(\rho,z)$ is therefore given by the difference $T(\rho,a)-T(\rho_0,a)$. In particular, we have
\begin{equation}
  \label{eq:beta_diff}
  \beta = T(a, a) - T(\rho,a)
\end{equation}
as the second Euler potential, wherein the $z$-dependence comes through $a=a(\rho,z)$.

Since the first term in Equation~(\ref{eq:beta_diff}) tends to $-\infty$, we temporarily evaluate it at a finite but undetermined position $(\rho_0,z_0)$ on the same streamline, which results in
\begin{align}
  \label{eq:hat_beta}
  & \hat{\beta}(\rho,z;z_0) := T \big( \rho_0, a(\rho,z) \big)
     -T \big( \rho, a(\rho,z) \big) \nonumber \\[0.20cm]
     &= \frac{1}{2} \ln \bigg( \frac{\sqrt{(r-z) \, (r + z - 1) + (z_0-1/2)^2}}{r-z} \bigg. \nonumber \\
     & \quad \bigg. -\frac{z_0-1/2}{r-z} \bigg) + z - z_0 \, ,
\end{align}
wherein the condition $a(\rho_0, z_0)=a(\rho,z)$ has been used to express $\rho_0$ as a function of $\rho$, $z$, and $z_0$.
Next, we compute $\nabla \hat{\beta}$, and only then let $z_0 \rightarrow \infty$, which results in
\begin{align}
  \label{eq:nabla_beta}
  \nabla \beta =& \lim_{z_0\rightarrow \infty} \nabla \hat{\beta}
  = \frac{\rho}{2 \, (r + z - 1) \, r} \, \vec{e}_\rho \nonumber \\[0.20cm]
  & + \left( 1 + \frac{r+z}{2 \, (r + z - 1) \, r} \right) \vec{e}_z \, .
\end{align}
Substituting this, together with $\nabla \alpha$ obtained from Equation~(\ref{eq:euler_alpha}), into Equation~(\ref{eq:euler}), the transversal component $\vec{B}_\mathrm{tr}$ is finally found to be
\begin{align}
  \label{eq:b_trans-1}
  \frac{(B_\mathrm{tr})_\rho}{B_\mathrm{tr,0}} =&\ \cos(\phi-\phi_0) \, \sqrt{\frac{r+z}{r+z-1}} \nonumber \\
   & \times \left(1 - \frac{r-z}{2 \, (r + z) \, r} \right) \\[0.20cm]
   \frac{(B_\mathrm{tr})_\phi}{B_\mathrm{tr,0}} =&\ - \sin(\phi-\phi_0) \, \sqrt{\frac{r+z}{r+z-1}} \\[0.20cm]
  \label{eq:b_trans-3}
  \frac{(B_\mathrm{tr})_z}{B_\mathrm{tr,0}} =&\ - \frac{\cos(\phi-\phi_0)}{2 \, r} \, \sqrt{\frac{r-z}{r+z-1}} \, .
\end{align}
The most general total solenoidal magnetic field solution to Equation~(\ref{eq:induct}), subject to the boundary conditions~(\ref{eq:Bo_rho})--(\ref{eq:Bo_z}), is then simply the sum of the respective longitudinal and transversal components~(\ref{eq:b_long}) and
(\ref{eq:b_trans-1})--(\ref{eq:b_trans-3}).
Various explicit coordinate representations for this total field are listed in Appendix~\ref{sec:appB}. \\

\subsection{General field structure}

\begin{figure*}
  \centering
  \includegraphics[width=0.98\textwidth]{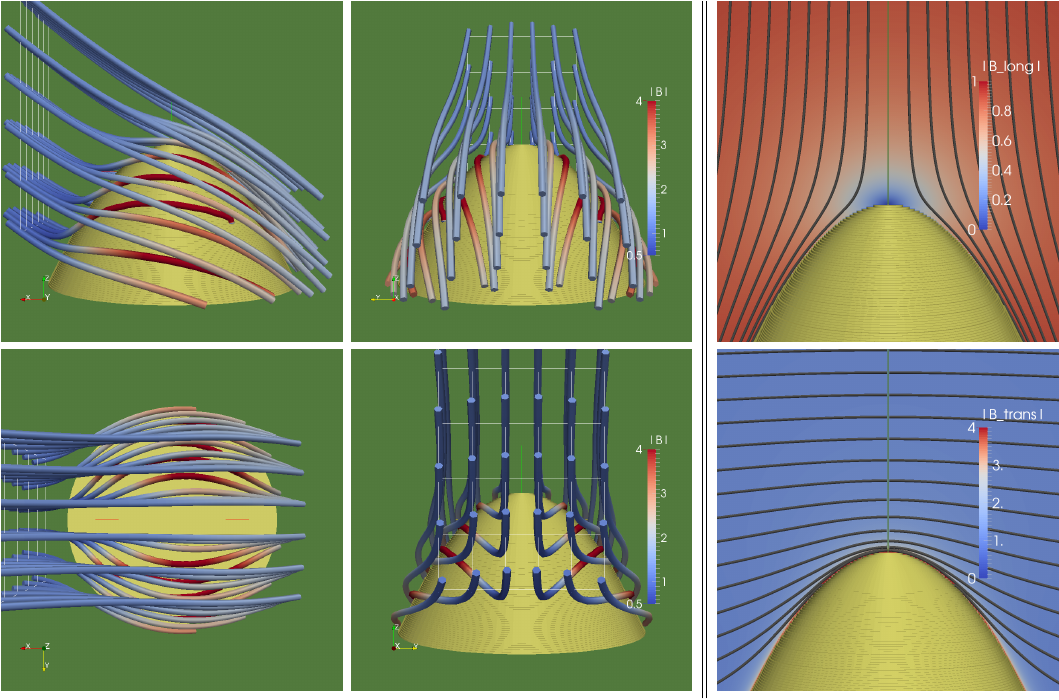}
  \caption{ \label{fig:fieldlines_3d}
  Left: four perspective renderings of field lines for boundary conditions $(B_{x 0}, B_{y 0}, B_{z 0}) = (1,0,1)$ draping around the obstacle (yellow) and colored according to $\|\vec{B}\|$. Since the latter tends to infinity near the obstacle's surface, contour values have been clamped within the interval $[0.5, 4.0]$.
  Right: field lines of the longitudinal (top) and transversal (bottom) parts of the magnetic field in the $(y=0)$ plane, with the background again color-coded to represent the field strength in this plane. The longitudinal field (whose field lines coincide with streamlines) vanishes at the stagnation point, while the transversal field (whose field lines coincide with isochrones) tends to infinity on the paraboloid's surface, but otherwise remains close to its upstream boundary value almost everywhere else. \\
  }
\end{figure*}

\noindent Figure~\ref{fig:fieldlines_3d} shows renderings of selected field lines from different vantage points. As expected, the field is almost homogeneous at large distances from the obstacle, but smoothly drapes around it in the vicinity of the latter. Along the $z$-axis, the respective magnitudes of the longitudinal and transversal components read
\begin{align}
  \big\| \vec{B}_\mathrm{lo}|_{\rho=0} \big\| &= \left(1- \frac{1}{2 \, z} \right) |B_{z 0}| \\[0.20cm]
  \big\| \vec{B}_\mathrm{tr}|_{\rho=0} \big\| &= \left(1- \frac{1}{2 \, z} \right)^{-1/2}
  \sqrt{ (B_{x 0})^2 + (B_{y 0})^2 } \, ,
\end{align}
implying that in the limit $z\rightarrow 1/2$ (i.e., at the stagnation point, located where the $z$-axis intersects the obstacle's surface), the longitudinal component vanishes while the transversal component tends to infinity. This behavior is analogous to what is found for the Rankine heliosphere, where the corresponding term in round brackets reads $(1-1/z^2)$ and likewise tends to zero as $z\rightarrow 1$ \citep[see Appendix~C of][]{Roeken_EA:2015}. Indeed, it can be verified that the resulting divergent field strength is not restricted to the stagnation point but extends to the entire surface of the paraboloid, also in exact analogy to the heliosphere case. The reason for this can be traced to a mismatch between the conflicting assumptions of stationarity and idealness, which enforces a perpetual pileup of magnetic flux that cannot be dissipated away. In a more realistic nonideal setting, the rate of incoming advected flux would be balanced by resistive dissipation, allowing for an equilibrium state with finite field strength. This, however, is beyond the scope of the present approach, which crucially relies on the frozen-in condition that is ensured by the ideal induction equation (\ref{eq:induct}).
A viable option to mitigate this problem for the heliopause, which would likely work also in the present case, has been presented by \citet{Florinski_EA:2024}. \\

\subsection{Isochrones revisited} \label{sec:isochrones}

\noindent For the applications to be discussed in Section~\ref{sec:apply}, it would be convenient to have at our disposal a set of isochrones that, while leading to the same magnetic field, are finite-valued in the entire region of interest. We will now introduce a procedure to arrive at such an isochrone field.

Our starting point is the ``quasi-isochrone'' (\ref{eq:hat_beta}), which only becomes a real isochrone in the limit $z_0 \rightarrow \infty$, passing through the point $(\rho,z)$.
The idea is then to exploit the gauge freedom of the Euler potential $\hat{\beta}$ by subtracting the fixed travel time of a fluid element traveling along the $z$-axis from $z=z_0$ to some other reference height $z=\za$. This reference travel time is conveniently calculated to be
\begin{align}
  \nonumber
  -\Delta\beta(\za; z_0) &:= -\int^\za_{z_0} \frac{\dd z}{u_z|_{\rho=0}}
  = -\int^\za_{z_0} \frac{\dd z}{(2 \, z)^{-1}-1} \\[0.20cm]
  &= \left[z + \frac{\ln(2 \, z - 1)}{2} \right]^\za_{z_0} ,
  \label{eq:delta_beta}
\end{align}
wherein we have used the fact that $\za,z_0>1/2$.
Since the choice of the value of $\za$ is arbitrary, we choose it such that the square bracket in Equation~(\ref{eq:delta_beta}) vanishes at $z=\za$, leading to the implicit equation
\begin{equation}
  \label{eq:za_gauge}
  2 \, \za + \ln (2 \, \za - 1 ) = 0 \, .
\end{equation}
The solution to this equation, which can be written as
\begin{equation}
  \za = \frac{W \left(e^{-1} \right)+1}{2} \approx 0.639
\end{equation}
using the Lambert~$W$~function, is immaterial for our purpose. What does matter is that it causes our renormalized Euler potential to assume the particularly simple form
\begin{align}
  \label{eq:beta_ren} \nonumber
  \betaren(\rho,z)
  &:= \lim_{z_0 \rightarrow \infty} \big[ \hat{\beta}(\rho,z;z_0) + \Delta\beta(\za; z_0) \big] \\[0.20cm]
  &= z+\frac{\ln(r+z-1)}{2} ,
\end{align}
a delightfully compact expression satisfying $\betaren(0,1)=1$ as a result of our choice~(\ref{eq:za_gauge}). It is straightforward to verify that $\nabla \betaren$ is equal to $\nabla \beta$ in Equation~(\ref{eq:nabla_beta}), which ensures that replacing $\beta$ by $\betaren$ in Equation~(\ref{eq:euler}) does not change the resulting magnetic field. \\

\section{Generalizations and applications}
\label{sec:apply}

\subsection{Different aspect ratios}
\label{sec:scaling}

\noindent The fact that our parabolic surface has a fixed aspect ratio may pose a problem for applications which require the obstacle's shape to be more elongated or compressed in the direction of its axis of symmetry. Suppose that one wishes to apply a linear scaling
\begin{equation}
    \label{eq:scaling_2d}
    (\rho,z) \rightarrow (\ell_\rho \, \rho, \ell_z \, z)
\end{equation}
with $\ell_{\rho}, \ell_z \in \mathbb{R}_{> 0}$, thus keeping the focal point fixed at the origin but moving the stagnation point from $z=1/2$ to $z=\ell_z/2$ and the cylindrical radius at which the paraboloid intersects the $(z=0)$ plane from $\rho=1$ to $\rho=\ell_\rho$. In an application, we might for instance imagine $\ell_{\rho}$ and $\ell_z$ to represent the object's actual spatial scales, expressed in problem-specific physical units of length.
Then, we may construct from the original $\vec{u}$ and $\vec{B}$ the scaled fields
\begin{align}
    \label{eq:u_sca}
    \vec{u}^{\, \mathrm{sc}}(\vec{r} \, ) :=&\ (\ell_\rho/\ell_z) \, u_\rho(\vec{r}_\ell) \, \vec{e}_\rho
    + u_z(\vec{r}_\ell) \, \vec{e}_z \\[0.20cm]
    \vec{B}_\mathrm{lo}\sca(\vec{r} \, ) :=&\ -B_{z 0} \,  \vec{u}^{\, \mathrm{sc}}(\vec{r}) \\[0.20cm]
    \vec{B}_\mathrm{tr}\sca(\vec{r} \, ) :=&\
    B_{\mathrm{tr}, \rho}(\vec{r}_\ell) \, \vec{e}_\rho +
    B_{\mathrm{tr}, \phi}(\vec{r}_\ell) \, \vec{e}_\phi \nonumber \\[0.10cm]
    \label{eq:Btr_sca}
    & + (\ell_z/\ell_\rho) \, B_{\mathrm{tr}, z} (\vec{r}_\ell) \, \vec{e}_z, 
\end{align}
in which the shorthands
\begin{align}
    \vec{r} &:= \rho \, \vec{e}_\rho + z \, \vec{e}_z \\[0.20cm]
    \label{eq:rL_cyl}
    \vec{r}_\ell &:= (\rho/\ell_\rho) \, \vec{e}_\rho + (z/\ell_z) \, \vec{e}_z
\end{align}
have been used as arguments. These fields $ \vec{u}^{\, \mathrm{sc}}$ and
$\vec{B}\sca = \vec{B}_\mathrm{lo}\sca + \vec{B}_\mathrm{tr}\sca$
can be shown to be (i) divergence-free, (ii) tangential to the scaled paraboloid, and (iii) an exact solution to the induction equation~(\ref{eq:induct}), subject to the same upstream boundary conditions as the original fields $\vec{u}$ and $\vec{B}$ prior to scaling. The last statement also follows from the more general proof given by \citet{Kleimann_EA:2016} in the context of so-called distortion flows, for which Equation~(\ref{eq:scaling_2d}) is a simple example.

Furthermore, given that said proof does not require any spatial symmetries, we may even take this idea one step further and replace the cylindrically symmetric scaling of Equation~(\ref{eq:scaling_2d}) with a still more general
\begin{equation}
    \label{eq:scaling_3d}
    (x,y,z) \rightarrow (\ell_x \, x, \ell_y \, y, \ell_z \, z) \, .
\end{equation}
This leads to what could be called a ``triaxial paraboloid,'' which intersects planes of constant $z$ in ellipses rather than circles, with the ellipse's semi-major axes at $z=0$ given by $\ell_x$ and $\ell_y$. The corresponding Cartesian flow field is then
\begin{equation}
    \label{eq:u_triaxial}
     \vec{u}^{\, \mathrm{sc}} = \frac{1}{2 \, r_\ell} \left[
    \frac{x \, \vec{e}_x + y \, \vec{e}_y}{r_\ell \, \ell_z +z} +\vec{e}_z
    \right] - \vec{e}_z \, ,
\end{equation}
wherein
\begin{equation}
    \label{eq:rL_car}
    r_\ell = \sqrt{(x/\ell_x)^2+(y/\ell_y)^2 +(z/\ell_z)^2}
\end{equation}
is now the triaxial analog to the absolute value of Equation~(\ref{eq:rL_cyl}). Explicit cylindrical and Cartesian components of $\vec{B}\sca$ for the respective scaling laws~(\ref{eq:scaling_2d}) and~(\ref{eq:scaling_3d}) are provided in Appendix~\ref{sec:appB} alongside the unscaled ones.

It seems worth noting that, while Equations~(\ref{eq:u_sca})--(\ref{eq:Btr_sca}) and also Equation~(\ref{eq:u_triaxial}) together with its associated magnetic field (see Equations~(\ref{eq:final_car_sc-1})--(\ref{eq:final_car_sc-3}) in Appendix \ref{sec:appB}) do constitute valid MHD solutions, the scaled velocity fields~(\ref{eq:u_sca}) and~(\ref{eq:u_triaxial}) are irrotational only for~$\ell_{\rho}=\ell_z$ and~$\ell_x=\ell_y=\ell_z$, respectively, and therefore in general do not have scalar potentials. The exact potential flows around scaled paraboloids could probably only be found be solving the Laplace equation~(\ref{eq:laplace}) for these geometries, which is beyond the scope of this paper. \\

\subsection{Advecting arbitrary magnetic fields}
\label{sec:general_b}

\noindent The choice of homogeneous boundary conditions (\ref{eq:Bo_rho})--(\ref{eq:Bo_z}) was  motivated not only as the simplest nontrivial option, but chiefly because typical astrophysical applications usually consider the incoming medium to be largely homogeneous. We may, however, generalize the procedure employed here to almost arbitrary magnetic fields, and study how they are deformed by the presence of the solid obstacle which they have to smoothly evade.

The underlying idea, which was already employed by \citet{Roeken_EA:2015}, is that streamlines (lines of constant $a$) and renormalized isochrones (lines of constant $\betaren$) as given by the respective Equations (\ref{eq:def_a}) and (\ref{eq:beta_ren}), together with the azimuthal coordinate $\phi$, can be interpreted as establishing a nonorthogonal coordinate system spanning the entire space exterior to the obstacle, and that, according to Cauchy's integral formalism \citep{Cauchy:1816}, the components of a field advected in the flow, when expressed in these coordinates, will remain constant throughout the entire transport.
As a consequence of the latter, we may think of the coordinates $(a, \phi, \betaren)$ as describing the physical situation in ``undistorted'' space, in which the obstacle is located at \mbox{$\betaren \rightarrow -\infty$}.
The transformation expressed in Equations~(\ref{eq:def_a}) and (\ref{eq:beta_ren}) then maps the usual cylindrical coordinates $(\rho,z) \in \mathbb{R}_{\ge 0} \times \mathbb{R}$ satisfying $2\, z \ge 1-\rho^2$ to coordinates $(a, \betaren) \in \mathbb{R}_{\ge 0} \times \mathbb{R}$, which are not restricted in any way as their coordinate surfaces ``warp'' around the obstacle. The conceptual distinction between these two sets of coordinates will be vital for the remainder of the paper.

A rather illustrative way of applying this mapping to the magnetic field, and similarly to other vector fields, is as follows. Suppose $\vec{B}_0(\rho,\phi,z)$ to be a given undistorted magnetic field, which may vary across and fill the entire space (thought of as being obstacle-free). 
This field generalizes the homogeneous field that would result from the static boundary conditions (\ref{eq:Bo_rho})--(\ref{eq:Bo_z}) being advected in a homogeneous, obstacle-free flow.
To obtain the corresponding field $\vec{B}(\rho,\phi,z)$ that results from the distortion of $\vec{B}_0(\rho,\phi,z)$ induced by the presence of the obstacle, it is useful to think of a vector as an actual ``arrow'' of infinitesimal length pointing from position $(\rho,\phi,z)$ to position $(\rho+\delta\rho, \phi+\delta\phi,z+\delta z)$, such that the components of the vectors $\vec{B}$ and $\vec{B}_0$ can be written as
\begin{align}
  \label{eq:Brhoz_deltas}
  \left( B_\rho, B_\phi, B_z \right)
  &= \left( \delta\rho, \rho \, \delta\phi, \delta z \right) \\[0.20cm]
  \label{eq:Babeta_deltas}
  \left( B_{0,a}, B_{0,\phi}, B_{0,{\betaren}} \right)
  &= \left( \delta a, a \, \delta\phi, \delta \betaren \right)
\end{align}
when expressed in regular and warped coordinates, respectively.
Without the obstacle, both coordinate systems, and therefore both magnetic fields, coincide exactly. To achieve this very desirable property was the motivation behind amending the second, rather than the first, Euler potential with a minus sign in Equation~(\ref{eq:def_beta}), since this ensures that
both $\nabla \betaren|_{\rho=0}$ and $\nabla z=\vec{e}_z$ are pointing into the same direction.
When the obstacle is present, however, the arguments $(\rho,z)$ of $\vec{B}_0$ are to be replaced by $(a,\betaren)$, which are themselves functions of $\rho$ and $z$.
Comparing the second components of Equations~(\ref{eq:Brhoz_deltas}) and (\ref{eq:Babeta_deltas}), we see that the toroidal component of $\vec{B}$ is therefore simply
\begin{equation}
  \label{eq:Bphi_trafo}
  B_\phi(\rho,\phi,z) = \frac{\rho}{a(\rho,z)}
  B_{0,\phi}\big(a(\rho,z), \phi, \betaren(\rho,z) \big) \, .
\end{equation}
For the computation of the poloidal components, however, we expand the first and third components of Equation~(\ref{eq:Babeta_deltas}) up to first order,
\begin{align}
  \label{eq:Babeta_from_Brhoz}
  \left( B_{0,a}, B_{0,\betaren} \right)
  &= \nonumber \left( \delta a, \delta\betaren \right) \\[0.20cm]
  &= \big( a(\rho+\delta\rho, z + \delta z) - a(\rho, z), \\
  &\quad \nonumber \hspace{0.25cm} \betaren( \rho+\delta\rho, z + \delta z) - \betaren( \rho, z) \big) \\[0.20cm]
  &= \nonumber \left( \PD{a}{\rho} \delta \rho + \PD{a}{z} \delta z ,
  \PD{\betaren}{\rho} \delta \rho + \PD{\betaren}{z} \delta z \right) .
\end{align}
The derivatives of $a(\rho,z)$ can be directly evaluated using the general relations~(\ref{eq:u_from_Psi}) and~(\ref{eq:rel_a-Psi}), yielding
\begin{align}
  \label{eq:da_drho}
  \PD{a}{\rho} &= \PD{~}{\rho} \sqrt{-2 \, \Psi} = - \frac{1}{\sqrt{- 2 \, \Psi}} \PD{\Psi}{\rho}
  = - \frac{1}{a} \, (\rho \, u_z) \\[0.20cm]
  \label{eq:da_dz}
  \PD{a}{z} &= \PD{~}{z} \sqrt{-2 \, \Psi} = - \frac{1}{\sqrt{- 2 \, \Psi}} \PD{\Psi}{z} = \frac{1}{a} \, (\rho \, u_\rho)
\end{align}
independently of the specific form of $a$ or $\Psi$ (and hence independently of the shape of the obstacle).
Then, inserting these expressions into Equation~(\ref{eq:Babeta_from_Brhoz}) and solving for the desired poloidal field components (\ref{eq:Brhoz_deltas}) results in
\begin{align}
  \label{eq:Brhoz_general}
  \colvec{B_\rho}{B_z} = \frac{1}{D} \left( \begin{array}{rc}
    -(a/\rho) (\partial_z \betaren) & \ \,\, u_\rho \\
    (a/\rho) (\partial_\rho \betaren) & \ \,\, u_z
  \end{array} \right) \colvec{B_{0,a}}{B_{0,\betaren}}
\end{align}
with the determinant
\begin{align}
  \label{eq:determinant} \nonumber
  D &:= \vec{u} \cdot \nabla \betaren
    = \PD{\rho}{t} \, \PD{\betaren}{\rho} + \PD{z}{t} \, \PD{\betaren}{z} \\
   &= \frac{\dd \betaren}{\dd t} = - 1 \, .
\end{align}
Here, the shape of the obstacle enters through the partial derivatives of $\betaren$, which are already known from Equation~(\ref{eq:nabla_beta}) because $\beta$ and $\betaren$ share the same gradient by construction.
Inserting these into Equation~(\ref{eq:Brhoz_general}) together with the components
(\ref{eq:u_rho_from_Psi}) and (\ref{eq:u_z_from_Psi}) of $\vec{u}$ and performing a series of algebraic simplifications, the explicit forms of the two poloidal components finally become
\begin{align}
  \label{eq:Brho_explicit} 
  B_\rho =&\ \sqrt{\frac{r+z}{r+z-1}} \left( 1 - \frac{r-z}{2 \, (r + z) \, r} \right) B_{0,a} \\
  & \nonumber - \frac{1}{2 \, r} \, \sqrt{\frac{r - z}{r + z}} \, B_{0,\betaren} \\[0.20cm]
  \label{eq:Bz_explicit} B_z =&\ - \frac{1}{2 \, r} \, \sqrt{\frac{r-z}{r+z-1}} \, B_{0,a} + \left(1- \frac{1}{2 \, r} \right) B_{0, \betaren} \, ,
\end{align}
wherein the arguments $a$ and $\betaren$ of $B_{0,a}$ and $B_{0,\betaren}$ are to be expressed through $\rho$ and $z$, analogous to the azimuthal component $B_{0,\phi}$ in Equation~(\ref{eq:Bphi_trafo}). Note that Equations~(\ref{eq:Brho_explicit}) and (\ref{eq:Bz_explicit}) are consistent with the formulas (\ref{eq:final_cyl-1}) and (\ref{eq:final_cyl-3}) found earlier for constant boundary conditions. \\

\subsection{A simple example: Cylindrical flux ropes}

\begin{figure*}
  \centering
  \includegraphics[width=\textwidth]{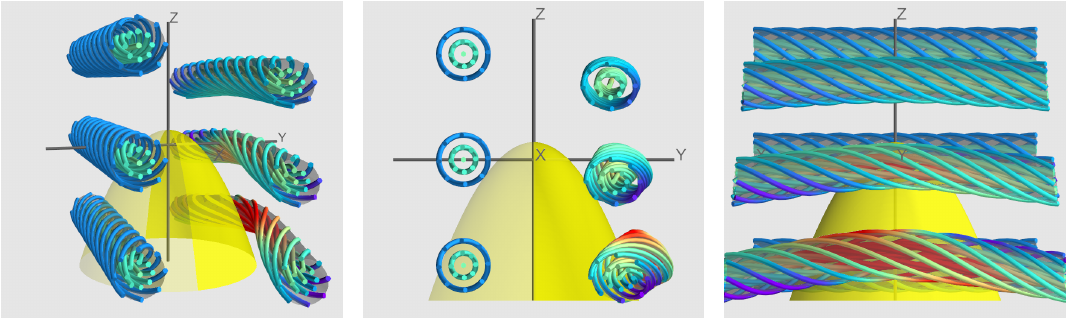}
  \caption{
    \label{fig:cool_render}
    Left: perspective rendering of several instances of the sample flux tube field (\ref{eq:sample_field}) with its axis parallel to the $x$-axis and passing through $y_\mathrm{c}=\pm 2$ and $z_\mathrm{c} \in \{-3,0,3 \}$, represented in its original form ($y<0$) vs.\ its shape when distorted and displaced by the presence of the parabolic obstacle ($y>0$). A semi-transparent cylinder of radius $0.8$ marks the outer boundary of the flux tube in each case. The color of field lines encodes the total field strength, increasing from blue to red. Middle: orthogonal projection of the same situation, viewed from the $+x$-direction. Right: again the same situation viewed from $+y$-direction. Note that, since this is also an orthogonal projection, the deformed flux tube segments in the front are indeed physically longer than those on the far side, rather than just appearing so because of their closer proximity to the observer. \\
  }
\end{figure*}

\noindent To illustrate the procedure outlined above using an astrophysically relevant example, let us consider the force-free magnetic field
\begin{equation}
\vec{B}_0 = \frac{\rho \, \vec{e}_{\phi} + \vec{e}_z}{1 + \rho^2} \, ,
\end{equation}
which has been used as a simple model for solar or interplanetary flux tubes 
\citep[e.g.,][]{Gold_Hoyle:1960,Kleimann_Hornig:2001,Aschwanden:2019}.
Its field lines form helical spirals of constant axial radius and constant thread lead $2\pi$. To make this example field more interesting, we reduce its symmetry by first rotating it by $90^\circ$ about the $y$-axis such that the $x$-axis becomes the new axis of symmetry, and then displace this axis of symmetry from $(y,z)=(0,0)$ to $(y_\mathrm{c},z_\mathrm{c})$. The result reads
\begin{align}
  \label{eq:sample_field}
  \nonumber \vec{B}_0 =&\ \bigl( \left[ \cos\phi - (z-z_\mathrm{c}) \sin\phi \right] \vec{e}_\rho \bigr. \\[0.15cm]
  &\ -\left[ \sin\phi +(z-z_\mathrm{c}) \cos\phi \right] \vec{e}_\phi + 
  \nonumber \big. \left[ \rho \sin\phi - y_\mathrm{c} \right] \vec{e}_z \bigr) \\[0.15cm]
  &\ \times \bigl[ 1 + (\rho \sin\phi-y_\mathrm{c})^2 + (z-z_\mathrm{c})^2 \bigr]^{-1} .
\end{align}
Applying Equations~(\ref{eq:Bphi_trafo}), (\ref{eq:Brho_explicit}), and (\ref{eq:Bz_explicit}) to this undistorted field then leads to the explicit form of the distorted field (not shown here for brevity).
A comparison of these two fields can be found in Figure~\ref{fig:cool_render}. 
In an astrophysical application, we might imagine the paraboloid to represent, for instance, a planetary magnetopause, while the sample field (\ref{eq:sample_field}) could model a solar/stellar magnetic flux tube that is incident upon the magnetosphere.

In such a situation, the travel-time interpretation of $-\betaren$ allows the deformed magnetic field to be viewed in two distinct ways:
\begin{enumerate}
\item According to the interpretation used in the derivation presented in Section~\ref{sec:general_b}, one may simply compare an undisturbed field to a version of itself that is disturbed by the presence of the (in this case) parabolic obstacle.
\item Since these fields are frozen into a stationary flow field, the situation can alternatively be viewed as a time-dependent advection in this flow. Therein, the undistorted field, which is advected in the homogeneous flow $\vec{u}_\mathrm{bg} = -\vec{e}_z$, satisfies
  \begin{equation}
    \vec{B}_0(\rho,\phi,z,t) = \vec{B}_0(\rho,\phi,z+\betaren, t-\betaren)
  \end{equation}
  and can thus be traced back to a time-dependent boundary condition at $z\rightarrow +\infty$. This property carries over to the distorted version of said field.
\end{enumerate}
As an illustration, Figure~\ref{fig:cool_render} shows three instances of the same flux rope at consecutive, uniformly spaced times (or, alternatively, a sequence of distinct, vertically displaced flux ropes at a single moment of time). \\

\subsection{Advecting arbitrary scalar fields}
\label{sec:general_sc}

\begin{figure*}
  \begin{center}
  \includegraphics[width=\textwidth]{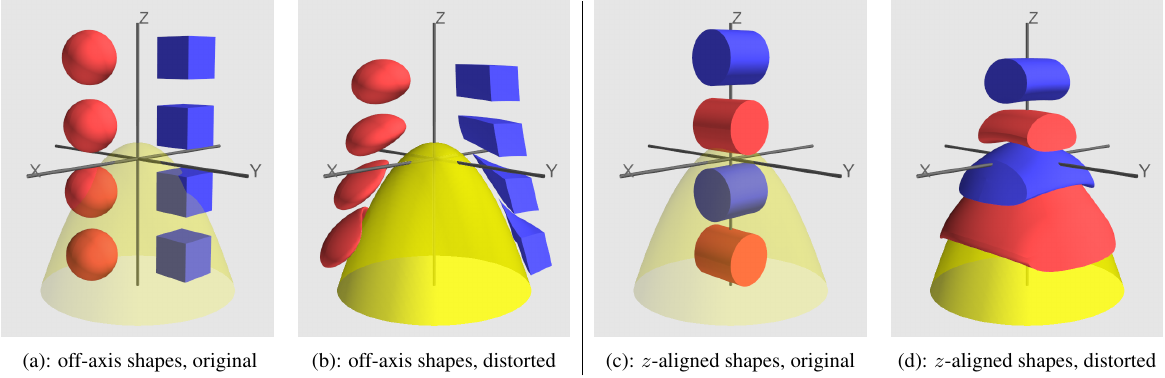}
  \caption{ \label{fig:deform_scalar}
    The deformation of basic geometric shapes induced by the presence of the parabolic obstacle. (a) Sequence of red spheres of radius $0.8$ centered at $(x_\mathrm{c}, y_\mathrm{c}, z_\mathrm{c})=(-1,-1,t_z)$ for $t_z \in \{-3,-1,1,3 \}$, together with a sequence of blue cubes of side length $1.2$ centered at $(x_\mathrm{c}, y_\mathrm{c}, z_\mathrm{c})=(1,1,t_z)$ and arranged parallel to, but not intersecting, the inflow axis. (b) The same shapes are severely distorted when passively advected in the flow, which is visualized by the pancake-like compression and flattening of these shapes as they are advected toward the obstacle, eventually sliding off it.
    (c) The same situation as in (a) but for a sequence of cylinders of height $1.6$ and radius $0.8$ alternating in color and orientation and placed right on the $z$-axis at $(x_\mathrm{c}, y_\mathrm{c}, z_\mathrm{c})=(0,0,t_z)$. (d) The cylinders are being compressed into (eventually) infinitesimally thin shells coating the paraboloid's surface. \\}
    \end{center}
\end{figure*}

\noindent The method outlined in Section~\ref{sec:general_b} may also be applied to the simpler case of scalar rather than vector fields. Given an undistorted scalar quantity $Q_0(\rho,\phi,z)$, its distorted counterpart simply reads
\begin{equation}
    Q(\rho,\phi,z)=Q_0\big( a(\rho,z), \phi, \betaren(\rho,z)\big) \, ,
\end{equation}
where $a(\rho,z)$ and $\betaren(\rho,z)$ are again provided by Equations~(\ref{eq:rel_a-Psi}) and (\ref{eq:beta_ren}). A sample visualization using spheres, cylinders with their axes of symmetry aligned with either the $x$- and $y$-direction, and cubes is shown in Figure~\ref{fig:deform_scalar}.
It is interesting to note that, due to the incompressibility of the underlying flow field, the total volume of all advected objects depicted in the figure is conserved at any time. (This would no longer be the case for the compressible flow discussed in Section~\ref{sec:compressible}, for which $\nabla \cdot \vec{u}\,\com \ne 0$.) The shapes used in this example might, for instance, stand in for dust clouds or regions of enhanced or decreased temperature in a stellar wind.

We call attention to the fact that the extension to magnetic and scalar fields with arbitrary spatial dependence only became possible through the renormalization of the second Euler potential, which is now directly associated with finite travel times. \\

\subsection{The inverse transformation}

\noindent To see more clearly how the presence of the parabolic obstacle displaces an advected fluid element, it would be convenient to know the explicit inverse $(a, \betaren) \mapsto (\rho, z)$ of the transformation given by Equations~(\ref{eq:rel_a-Psi}) and (\ref{eq:beta_ren}). Unfortunately, the implicit nature of the latter equation forestalls the derivation of analytical formulas to this effect, so here at last, a numerical procedure becomes mandatory. We will now present and discuss such a procedure.

The first transformation formula (\ref{eq:rel_a-Psi}) can be easily solved for $\rho$ to yield
\begin{equation}
  \label{eq:rho_a}
  \rho = \sqrt{ a^2- (z-1/2) + \waz }
\end{equation}
with the shorthand
\begin{equation}
  \waz := \sqrt{a^2 + (z-1/2)^2} \ ,
\end{equation}
which will be of further use shortly.
We then insert Equation~(\ref{eq:rho_a}) into the second transformation formula (\ref{eq:beta_ren}) to obtain
\begin{equation}
  \label{eq:beta_z}
 \betaren = z + \frac{\ln \big( \waz + z-1/2 \big)}{2}  =: \beta_a(z)
\end{equation}
as an implicit equation for the coordinate $z$. For fixed $a$, this equation only depends on a single argument and may be interpreted as giving the isochrone label of a fluid element located at height $z$ while sliding along streamline $a$.
It is clear from this interpretation, and also easy to verify via
\begin{equation}
  \beta_a\p(z) \equiv \frac{\dd \beta_a(z)}{\dd z} = 1 + \frac{1}{2 \, \waz} > 0 \, ,
\end{equation}
that $z \mapsto \beta_a(z)$ is strictly monotonous. Since it is also strictly concave, i.e., $\beta_a^{\prime\prime}(z)<0$ for $z>1/2$, Equation~(\ref{eq:beta_z}) is well suited for an iterative solver based on Newton's method, which in this case reads
\begin{align}
  z_{k+1} &= z_k - \frac{\beta_a(z_k) - \betaren}{\beta_a\p(z_k)} \\[0.20cm]
  \nonumber &= \frac{z_k + w_a(z_k) \left[ 2 \, \betaren - \ln \big( w_a(z_k) +z_k -1/2 \big) \right] }{1+2 \, w_a(z_k)} \ .
\end{align}
For a given starting value, which we here choose to place at $z_1=1$, the solver quickly converges to the desired solution $z = \lim_{k \rightarrow \infty} z_k$ of Equation~(\ref{eq:beta_z}).

For the special case $a=0$, we need to bear in mind that, since this is the only streamline that does not extend indefinitely into the negative $z$~direction but rather ends at the stagnation point, it may happen that $z_{k+1} \le 1/2$ for some $k$, indicating a position on or inside the paraboloid. A possible strategy, which we shall adopt here without examining further options for its optimization, is to replace $z_{k+1}$ by
$(1/2 + z_k)/2$, i.e., the midpoint between the stagnation point and the previous approximate value.
This ensures that $z_{k+1}$ is both a valid exterior position and closer to the desired result than $z_k$ had been.

Since Newton's method converges quadratically, satisfactory results are typically obtained within just a few iterations. Tentative computations for target values on a grid of $(a, \betaren) \in [0,4] \times [-8,4]$ with a spacing of $0.1$ in both directions show that it takes between $N=2$ and $7$ iterations ($3.34$ on average) to warrant that $| \beta_a(z_{1+N})-\betaren | <10^{-9}$ if $a>0$.
For $a=0$, however, approximately 13 iterations are typically needed (except for $\betaren=1$, where $z=1$ by construction). \\

\section{Extension to compressible flow}
\label{sec:compressible}

\noindent \citet{Kleimann_EA:2017} presented a generalization of the \citet{Roeken_EA:2015} MHD model of the outer heliosheath to compressible flow, thus allowing the number density $n$ to vary in a physically more realistic way. We will now derive the corresponding case for the flow around the paraboloid. \\

\subsection{Density and flow fields}

\noindent The key idea behind said generalization is to reinterpret the potential $\Phi$ as one for momentum density $n \, \vec{u}$ rather than $\vec{u}$, thereby ensuring that the flow's streamline structure, and in particular the shape of the obstacle, remains unchanged while permitting $n$ to vary along streamlines. This approach has the appealing feature that Equation~(\ref{eq:laplace}) now becomes 
\begin{equation}
  \label{eq:continuity}
  \nabla\cdot \left(n \, \vec{u} \right)=0 \, ,
\end{equation}
implying that the compressible extension of our model continues to conserve mass by construction. We handle the newly introduced degree of freedom by additionally requiring the projection of the momentum balance equation
\begin{equation}
  \label{eq:momentum}
  m^2 \, n \, (\vec{u} \cdot \nabla) \vec{u} = - \frac{\nabla P}{\gamma} = - n^{\gamma-1} \, \nabla n
\end{equation}
upon streamlines for a fluid whose gas pressure $P$ obeys the adiabatic equation of state
\begin{equation}
  \label{eq:adiabatic_eos}
  P = n^\gamma
\end{equation}
to be satisfied. Here, $\gamma$ is the adiabatic index and the new parameter $m \in [0,1]$ denotes the sonic Mach number at upstream infinity, with $P$ consistently being normalized to its respective value at the upstream boundary. 
{We note that the procedure to be described is also applicable to the more general equation of state $P = p_0 + p_1 \, n^\gamma$ with constants $p_0$ and $p_1>0$, which has been suggested in the context of suprathermal particle populations, particularly those described by $\kappa$-distributions \citep[e.g.,][and references therein]{Silveira_EA:2021, Lazar_EA:2020}. In that case, the resulting density pattern does not depend on the ``vacuum pressure'' contribution $p_0$ because it is only the gradient of $P$ that enters Equation~(\ref{eq:momentum}). The prefactor $p_1$, on the other hand, can conveniently be absorbed into a redefined Mach number $m^\prime=m/\sqrt{p_1}$. We further note that our derivation is not conditional on the interpretation of $\gamma$ as the fluid's adiabatic index, but is just as valid for other constant exponents \citep[e.g.,][]{Scudder:1992, Meyer-Vernet_EA:1995}. For these reasons, we will continue to use the Maxwellian adiabatic equation of state~(\ref{eq:adiabatic_eos}) in the ensuing considerations.}

Fortunately, most of the construction presented in \citet{Kleimann_EA:2017} does not make any reference to a particular flow field, and is thus also applicable to the situation at hand without any modifications. In particular, the desired density field $n$ may be obtained from the same implicit equations, namely
\begin{equation}
  \label{eq:n-A_adiab}
  m^2 \A = n^2
  \left[m^2 - \frac{2 \, (n^{\gamma-1}-1)}{\gamma-1} \right]
\end{equation}
for $\gamma \ne 1$ and
\begin{equation}
  \label{eq:n-A_isoth}
   m^2 \A = n^2 \left[m^2 - 2 \, \ln (n) \right]
\end{equation}
for the isothermal case $\gamma = 1$.
Here, $\A$ is a
scalar function of meridional position $(\rho,z)$, which, for the present paraboloid case, evaluates to
\begin{equation}
  \A := (\nabla \Phi)^2
  = 1 - \frac{1}{r} + \frac{1}{2 \, (r+z) \, r} \, .
\end{equation}
This function maps the entire region exterior to the obstacle onto the interval $[0,1)$, with the lower boundary $0$ and the limiting case $1$ being exclusively attained at the stagnation point and at upstream infinity, respectively. Note that both Equations~(\ref{eq:n-A_adiab}) and (\ref{eq:n-A_isoth}) are consistent with the uniform density case $n = 1$ being recovered for $m = 0$.

Continuing the analogy to \citet{Kleimann_EA:2017}, Equation~(\ref{eq:n-A_isoth}) for the isothermal case can be ``solved'' using the principal branch $W_0$ of the Lambert~$W$~function as
\begin{equation}
  \label{eq:imp_gamma_eq1}
  n(\A) = \exp \left( \frac{m^2}{2}
    + \frac{1}{2} \, W_0 \left(- m^2 \exp(- m^2) \, \A
    \right) \right) ,
\end{equation}
which confirms that the peak density is reached at the stagnation point and amounts to $n(0) = \exp(m^2/2) \le \exp(1/2) \approx 1.649$.

For the mono-atomic ideal gas case $\gamma=5/3$, upon which we shall specialize in the following besides $\gamma=1$, solving Equation~(\ref{eq:n-A_adiab}) is equivalent to finding the two real-valued roots of the fourth-order polynomial
\begin{equation}
  \label{eq:imp_gamma_ne1}
  \xi^4 - \left(1+m^2/3\right) \, \xi^3 + \left(m^2/3\right) \A \, ,
\end{equation}
where $\xi:= n^{2/3}$. The smaller one of these roots, which is always below unity, corresponds to an unrealistic density field that attains its (global) minimum, rather than its maximum, at the stagnation point. This behavior can be observed in the isothermal case as well, where it is induced by using the $W_{-1}$ branch of the Lambert~$W$~function.
Moreover, the physical significance of a density field that is based on the smaller root is further called into question by the fact that the corresponding flow field is supersonic everywhere, that the density vanishes at the stagnation point for any $m$, and that setting $m=0$ fails to recover the incompressible limit ($n=1$ everywhere) but rather describes a ``vacuum solution'' ($n=0$ everywhere).
For these reasons, we consider the larger root, which is always above unity and does not exhibit any of the above-mentioned issues, to be the physically relevant one in the present context. 

Although the roots of the quartic (\ref{eq:imp_gamma_ne1}) can in principle be found using Ferrari's formula, a more convenient expression would be desirable, and the same is certainly true for Equation~(\ref{eq:imp_gamma_eq1}) involving the Lambert~$W$~function.
For this reason, we again repeat the same steps already applied in \citet{Kleimann_EA:2017}, namely, to replace both Equation~(\ref{eq:imp_gamma_eq1}) and $n(\A)$ in the exact (but implicit) expression~(\ref{eq:imp_gamma_ne1}) by the second-order polynomial $n_2(\A)$ whose coefficients are adjusted such that it coincides with $n(\A)$ at $\A=0$ (the stagnation point) and $\A=1$ (infinity), and also has the same derivative at $\A=0$. As a simpler, less accurate alternative, we also consider the first-order polynomial $n_1(\A)$, which only fulfills the first two conditions. Reproducing Equations~(44)--(47) from \citet{Kleimann_EA:2017}, we get
\begin{align}
  \label{eq:nb_approx1}
  n_1(\A) &= \nsp + (1-\nsp) \, \A \\[0.20cm]
  \label{eq:nb_approx2}
  n_2(\A) &= \nsp + D_0 \, \A + (1-\nsp-D_0) \, \A^2
\end{align}
with
\begin{equation}
  \label{eq:n_max}
  n_{\rm sp} := n(0) = \left\{ \begin{array}{lcl}
      \exp \left(m^2/2 \right)     &\ : \ & \gamma=1 \\[0.20cm]
      \left( 1+m^2/3 \right)^{3/2} &\ : \ & \gamma=5/3
    \end{array} \right. 
\end{equation}
the stagnation point density and
\begin{equation}
  \label{eq:D_0}
  D_0 := -\frac{m^2}{2} \times
  \left\{ \begin{array}{lcl}
      \exp \left(-m^2/2 \right) &\ : \ & \gamma=1 \\[0.20cm]  \displaystyle
      \left( 1+m^2/3 \right)^{-5/2}
      &\ : \ & \gamma=5/3
    \end{array} \right.
\end{equation}
the derivative of $n(\A)$ at $\A=0$.

\begin{figure*}
  \begin{center}
  \includegraphics[width=\textwidth]{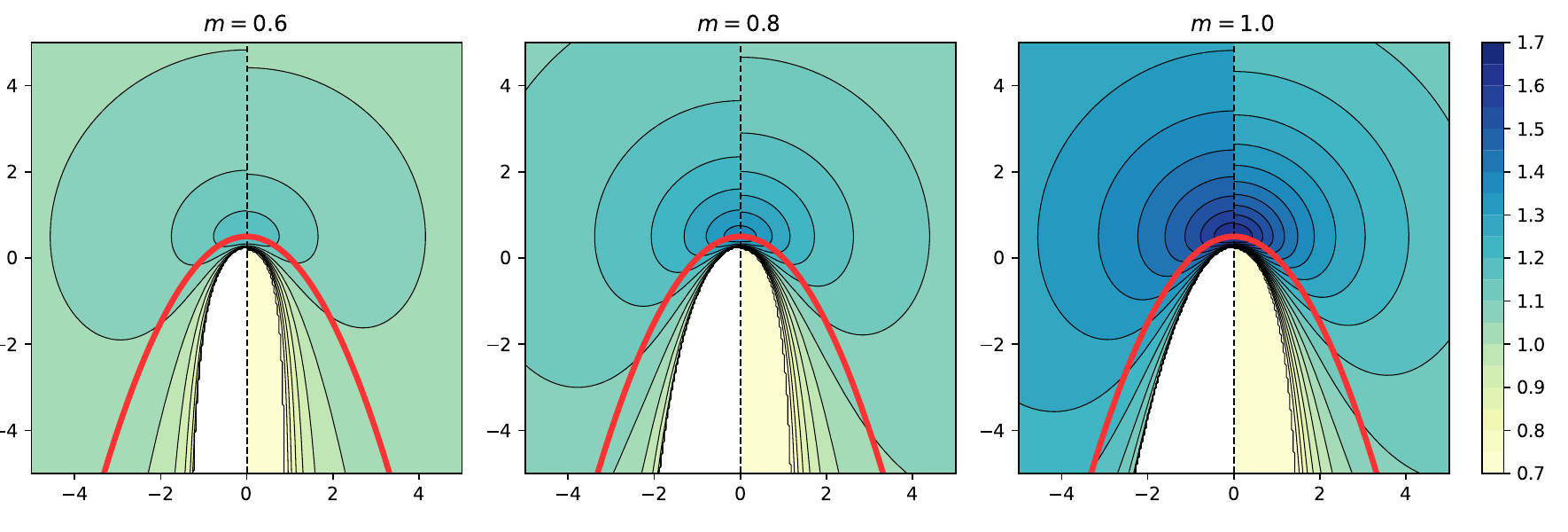}
  \end{center}
  \caption{ \label{fig:density}
    Density contours for three different values of the upstream Mach number $m$, in each case plotted using both the exact profile (Equation~(\ref{eq:imp_gamma_eq1}), left side) vs.\ the second-order polynomial approximation (Equation~(\ref{eq:nb_approx2}), right side). The red curves mark the parabolic obstacle.
    Density values below 0.7 are not shown, and no exact solution exists in the white region. \\
  }
\end{figure*}

Figure~\ref{fig:density} shows isothermal density contours for selected values of $m$, in each case comparing the exact formula~(\ref{eq:imp_gamma_eq1}) with its approximation~(\ref{eq:nb_approx2}). Both exhibit a very similar overall behavior, and, by construction, agree exactly at the stagnation point and at infinity.
As can also be seen from the figure, the density at any point exterior to the obstacle is never below the density of incompressible flow. This constitutes a remarkable departure from the heliosphere case \citep{Kleimann_EA:2017}, which features extended regions in the flanks where the flow speed is higher, and the density correspondingly lower, than in the incompressible situation. The density contours of the adiabatic case are very similar to those of the isothermal case, and are not shown for exactly this reason.

It is worth noting that, while the polynomial approximation can be evaluated at any position $(\rho,z)$, the exact expression exhibits a region in which it becomes complex-valued and hence meaningless (indicated in white in Figure~\ref{fig:density}). In the heliosphere case, this region grows as $m$ increases, eventually engulfing the entire downwind half-space as $m \rightarrow 1$. This, however, is apparently different for the paraboloid, where it remains confined to the interior of the obstacle, thereby allowing the exact expression (\ref{eq:imp_gamma_eq1}) to be used for any $m \le 1$ (and technically even beyond that, though of course no bow shock can be expected) if so desired. \\

\subsection{The magnetic field for compressible flow}

\noindent Because the magnetic field previously derived from Equation~(\ref{eq:induct}) is frozen into $\vec{u}$ rather than $n \, \vec{u}$, the question arises as to how the transition from incompressible to compressible flow changes the magnetic field. This question was already investigated by \citet{Kleimann_EA:2017} for the Rankine-type heliosphere, resulting again in fully analytical, albeit more involved, formulas for the compressible magnetic field components. Rather than applying their procedure to the present parabolic case, we approach the problem using the framework of Euler potentials, as was already done for the incompressible case in Section~\ref{sec:magnetic}.

We begin by deriving the longitudinal component, which is parallel to $\vec{u}\,\com$, and may therefore be written as $\vec{B}_\mathrm{lo}\com = \omega \, \vec{u}\,\com$, where $\omega$ is a yet undefined scalar function of position. From
\begin{align} \nonumber
  0 =&\ \nabla\cdot \vec{B}_\mathrm{lo}\com = \nabla\cdot [ \omega \, \vec{u}\inc/n ] \\[0.20cm] 
  =&\ (\omega/n) \, \underbrace{\nabla\cdot \vec{u}\inc}_{=0}
  +  \left( \vec{u}\inc \cdot \nabla \right) (\omega/n)
\end{align}
we can conclude that $\omega/n$ is constant along any streamline, though it may still vary across streamlines. However, consistency with the upstream boundary condition can only be ensured if $\omega/n$ equals the global constant $-B_{z 0}$. Therefore,
\mbox{$\vec{B}_\mathrm{lo}\com = -B_{z 0} \, \vec{u}\inc = \vec{B}_\mathrm{lo}\inc$}, implying that the longitudinal component remains completely unaffected by the transition to compressibility.

For the transversal component, we use the same first Euler potential
$\alpha = B_\mathrm{tr,0}\, a\, \sin(\phi-\phi_0) = -a \, B_{\phi 0}$ as in the incompressible case, whose gradient evaluates to
\begin{equation}
  \nabla \alpha = \frac{\rho}{a} \, B_{\phi 0} \, \bigl(u_z\inc \, \vec{e}_\rho - u_\rho\inc \, \vec{e}_z\bigr) + \frac{a}{\rho} \, B_{\rho 0} \, \vec{e}_\phi  
\end{equation}
according to Equations~(\ref{eq:da_drho}) and (\ref{eq:da_dz}). Similarly to the corresponding incompressible expression in Section \ref{Sec-3-2}, the gradient
of the second Euler potential becomes
\begin{align} \nonumber
    \nabla \beta\com =&\ \lim_{z_0\rightarrow \infty} \nabla \hat{\beta}\com = - \lim_{z_0\rightarrow \infty} \nabla \int_{\rho_0}^\rho \frac{\dd\rho\p}{u_\rho\com} \\[0.20cm]
    =&\ - \lim_{z_0\rightarrow \infty} \nabla \int_{\rho_0}^\rho \frac{\nsp + \nu_1 \A + \nu_2 \A^2}{u_\rho\inc} \, \dd\rho\p \nonumber
    \\[0.20cm] 
    =&\ \nsp \, \nabla \beta\inc - \nu_1 \, \vec{K}_1(\rho, z) - \nu_2 \, \vec{K}_2(\rho, z) \, ,
\end{align}
with $\rho_0 =\rho_0(\rho, z; z_0)$ as given in Equation~(\ref{eq:rho0}), the functions
\begin{equation} \label{Ks}
  \vec{K}_k(\rho, z) := \lim_{z_0\rightarrow \infty} \nabla \int_{\rho_0}^\rho \frac{\A^k}{u_\rho\inc} \, \dd\rho\p \, , 
\end{equation}
$k \in \{1, 2\}$, explicitly specified in Equations~(\ref{eq:K1}) and~(\ref{eq:K2}), and the constants $\nu_{1,2}$ to be read off Equation~(\ref{eq:nb_approx1}) or~(\ref{eq:nb_approx2}). 
With the above, we may express the magnetic field 
\begin{equation}
\vec B^\mathrm{C} = \vec{B}_\mathrm{tr}\com + \vec{B}_\mathrm{lo}\com = \nabla \alpha \times \nabla \beta\com - B_{z 0} \, \vec{u}\inc
\end{equation}
for compressible flow in terms of the magnetic field $\vec B\inc$ for incompressible flow given in Equations~(\ref{eq:Brho_explicit}) and (\ref{eq:Bz_explicit}). Fortunately, the toroidal part $B_\phi\com$ of the compressible magnetic field can be related to the respective incompressible component $B_\phi\inc$ in a way that is independent of the specific forms of the poloidal flow field and the gradient of the compressible second Euler potential. This is evident from the short computation
\begin{align}
  \label{eq:Bcomp_phi} \nonumber
  B_\phi\com &= B_{\phi, \mathrm{tr}}\com + B_{\phi, \mathrm{lo}}\com
  = \left[\nabla\alpha \times \nabla \beta\com \right]_\phi + 0 \\[0.20cm]
  &= -\underbrace{(\rho/a) B_{\phi 0}}_{=B_\phi\inc} \, n \, 
   \underbrace{\vec{u}\com \cdot \nabla \beta\com}_{= - 1} 
  = n \, B_{\phi}\inc \, ,
\end{align}
which uses the compressible version of the determinantial relation~(\ref{eq:determinant}). For the poloidal parts $B_{\rho}\com$ and $B_z\com$, however, the explicit forms of both the flow field and the gradient of the compressible second Euler potential become relevant. Accordingly, we obtain 
\begin{align}
  \label{eq:Bcomp_rho}
  B_{\rho}\com &= \nsp \, B_{\rho}\inc - \mathcal{Z}_z + (\nsp - 1) \, u_{\rho} \, B_{z 0} \\[0.20cm]
  \label{eq:Bcomp_z}
  B_z\com &= \nsp \, B_z\inc + \mathcal{Z}_{\rho} + (\nsp - 1) \, u_z \, B_{z 0}\, ,
\end{align}
where 
\begin{equation}
  \label{eq:Z}
  \vec{\mathcal{Z}} := \frac{a}{\rho} \, \bigl(\nu_1 \, \vec{K}_1 + \nu_2 \, \vec{K}_2\bigr) \, B_{\rho 0} \, .
\end{equation}

We remark that, despite being well-defined at any position outside the paraboloid, the function $\vec{K}_2$ is somewhat cumbersome to numerically evaluate in the vicinity of the $(a=0)$ streamline (see Equation~(\ref{eq:K2}) in Appendix~\ref{sec:appC}). Therefore, we now directly use the induction equation to compute expressions for the compressible field components on the $z$-axis that are both well-defined and easy to evaluate, yielding
\begin{align}
  \label{eq:Bcomp_xy_axis}
  \frac{B_x\com|_{\rho=0}}{B_{x0}} &= n_{1,2}|_{\rho=0} \, \biggl(1 - \frac{1}{2 \, z}\biggr)^{- 1/2} = \frac{B_y\com|_{\rho=0}}{B_{y0}} \\[0.20cm]
  \label{eq:Bcomp_z_axis}
  \frac{B_z\com|_{\rho=0}}{B_{z0}} &= 1 - \frac{1}{2 \, z}
\end{align}
in analogy to \citet{Kleimann_EA:2017}.
These expressions are of course compatible with the magnetic field components (\ref{eq:Bcomp_phi})--(\ref{eq:Bcomp_z}) evaluated at $\rho = 0$. Moreover, in the limit $z \rightarrow \infty$ the latter components tend to the boundary field $\vec{B}_0$ at upstream infinity.
It also seems worth noting that our compressible results are ``approximative'' only in the sense that $n_{1,2}$ do not exactly fulfill the momentum balance equation~(\ref{eq:momentum}). The magnetic field thus derived is nonetheless an exact solution to the induction equation~(\ref{eq:induct}) for $\vec u\,\com = \vec u\inc/n_{1,2}$.

\begin{figure}
    \includegraphics[width=0.48\textwidth]{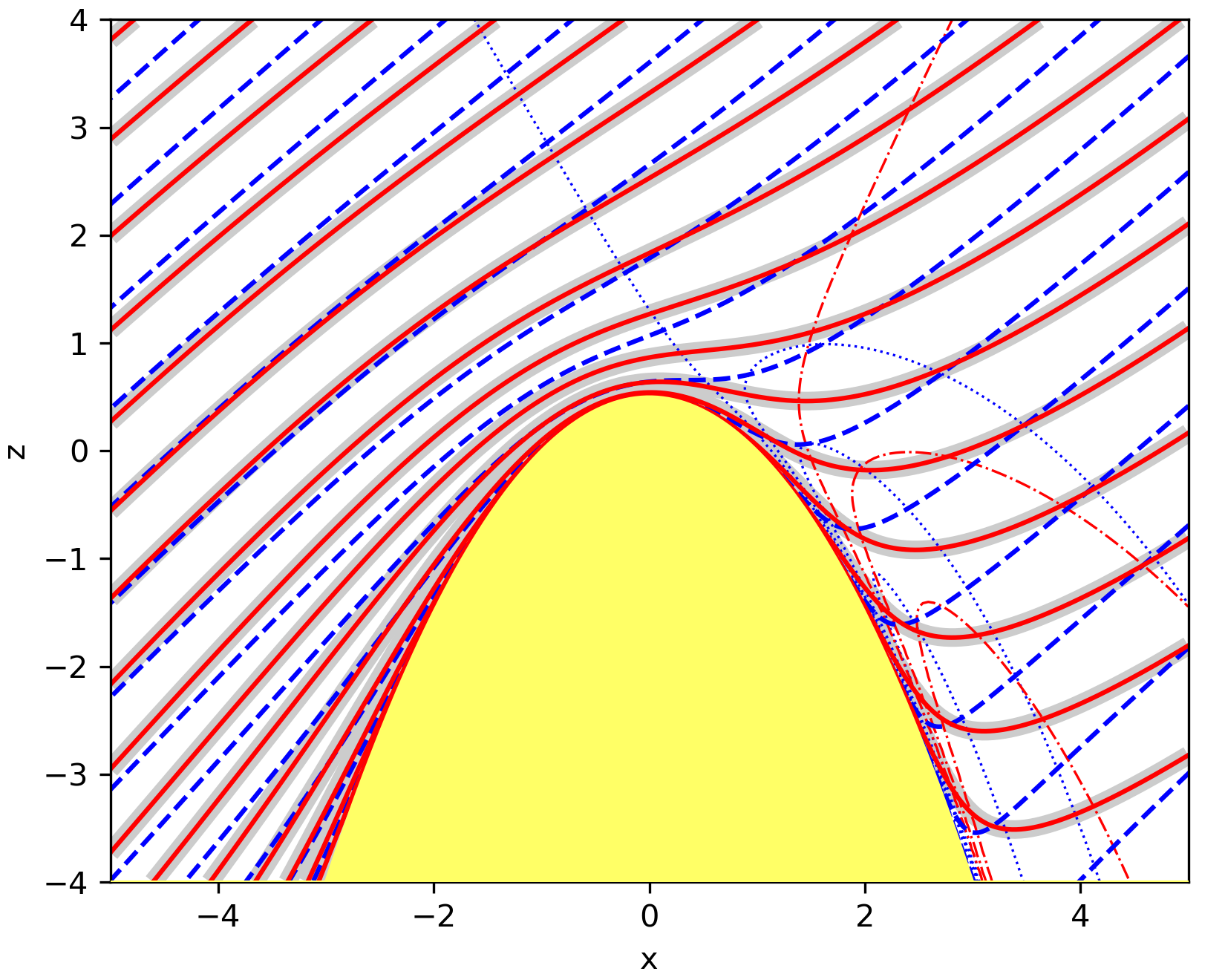}
    \caption{ \label{fig:fieldlines_2d}
    Field lines of $\vec{B}\com$ (compressible, Equations~(\ref{eq:Bcomp_rho})--(\ref{eq:Bcomp_z}), $m=1$, $\gamma=1$, solid, red) for the first-order density approximation~(\ref{eq:nb_approx1}), contrasted with those of $\vec{B}\inc$ (incompressible, Equations~(\ref{eq:final_cyl-1})--(\ref{eq:final_cyl-3}), $m=0$, dashed, blue) in the $(y=0)$~plane. The former are amended with the corresponding contours of $\beta\com_\mathrm{ren}$ (thick gray lines), and both clearly coincide.
    Additionally, magnetic traps are indicated by the thin red and blue contour lines, marking the regions at which the respective field strengths of $\vec{B}\com$ (red, dash-dotted) and $\vec{B}\inc$ (blue, dotted) have decreased to $[100, 80, 60, 40]$\% of their common boundary value of $\|\vec{B}_0\com\| = \|\vec{B}_0\inc\| = \sqrt{2}$.}
\end{figure}

Figure~\ref{fig:fieldlines_2d} compares the compressible field to its incompressible counterpart. While both drape around the obstacle in a qualitatively similar manner, it can be seen that the compressible field not only tends to be less sharply curved in the vicinity of the obstacle, but also starts to deviate from straight lines already at larger upwind distances, which is consistent with the general behavior found in the compressible heliospheric case.
Regarding the total field strength, we note in particular the occurrence of \textit{magnetic traps} of $\vec{B}^\mathrm{I,C}$, regions of markedly reduced field strength in which charged particles may gyrate for extended periods of time before they can escape again. (See \citet{Florinski_EA:2024} for an investigation of such traps for the heliospheric case based on an extension of the \citet{Roeken_EA:2015} field model.)

In order to extend the scalar advection processes discussed in Section~\ref{sec:general_sc} to the compressible case, a correspondingly modified renormalized second Euler potential $\beta_\mathrm{ren}\com$ is required. The explicit form of the latter, which is derived in Appendix~\ref{sec:appD}, allows applications such as those presented in Figure~\ref{fig:deform_scalar} to be obtained also for compressible flow with relative ease.
As a further example of its usefulness, we observe that, for the constant Cartesian components $(B_{x 0}, 0, B_{z 0})$ of a planar homogeneous field, the general field line equation
\begin{equation}
    \frac{\dd x}{B_x} = \frac{\dd z}{B_z}
\end{equation}
can be trivially integrated to the condition
\begin{equation}
    B_{z 0} \, x - B_{x 0} \, z = \mathrm{const.}
\end{equation}
describing the associated straight field lines. However, since the transformations $(\rho,z) \leftrightarrow \bigl(a(\rho,z), \betaren(\rho,z) \bigr)$ employed in Section~\ref{sec:general_b} continue to be valid also for the compressible case, field lines of the latter can easily be drawn as contours of 
\begin{equation}
    \label{eq:fieldline_contours}
    B_{z 0}  \ a \Big(\frac{x}{\cos\phi},z \Big) \, \cos\phi
    -B_{x 0} \ \beta\com_\mathrm{ren} \Big(\frac{x}{\cos\phi},z \Big) \, ,
\end{equation}
where $x=\rho \, \cos\phi$, $\phi \in \{0,\pi \}$, is used rather than $\rho$ to be able to access the negative values of the $x$ coordinate.
This method allows us to draw field lines without the otherwise mandatory recourse to a numerical field line tracer, as exemplified in Figure~\ref{fig:fieldlines_2d}. Therein, contours of expression~(\ref{eq:fieldline_contours}) are drawn using the compressible second Euler potential~(\ref{eq:beta_ren_C}), evaluated for the first-order density approximation~(\ref{eq:nb_approx1}), i.e., with $\nu_1=1-\nsp$ and $\nu_2=0$. With the corresponding field lines (derived by traditional numerical integration) overplotted, and both adjusted to pass through the same set of seed points, the excellent agreement of the results from both methods is immediately evident.

Although in this plot the first-order approximation is preferred over its second-order counterpart for technical reasons (detailed in Appendix~\ref{sec:appC}), we see that even this simpler first-order variant provides a realistic and physically plausible model of a magnetic field draping around a parabolic obstacle immersed in a compressible flow. \\

\section{Summary and conclusions}
\label{sec:summary}

\noindent In this work, we presented a (magneto)hydrodynamical model for laminar, inviscid flow ahead of and around a solid obstacle of paraboloid form whose axis of symmetry coincides with that of the incident flow. For the case of the fluid being both incompressible and irrotational, we derived compact formulas for the scalar flow potential and the stream function, and hence for the velocity field. This velocity field was then used as an input to the induction equation of ideal magnetohydrodynamics, which we solved for the magnetic field subject to a homogeneous upstream boundary condition of arbitrary inclination. The resulting field components can be written in compact and explicit form, involving neither unevaluated integrals nor special functions. This model constitutes the first (and therefore so far also the only) known nontrivial axially symmetric magnetohydrodynamical solution for obstacle-modulated flow with this property.

Moreover, we used a renormalization procedure to address the problem of the infinite travel time spent by a fluid element moving from upstream infinity to the vicinity of the obstacle. As a further result of this renormalization procedure, a simple one-to-one mapping between distorted and undistorted coordinate space (where in the latter the obstacle located at downstream infinity) is established that not only eases visualization, but also allows to describe the profound deformations experienced by arbitrary scalar or vector fields through the presence of the obstacle around which they are advected. In one direction, this mapping is available as an explicit formula, while in the opposite direction, recourse to a (quickly converging) numerical algorithm becomes mandatory. When interpreting the obstacle's surface as the magnetopause of a planet in a subsonic stellar wind, typical astrophysical applications include the modeling of transients, such as magnetic clouds or local density enhancements, embedded in the flow that is incident on the magnetosphere.

In a further step, we generalized the above model first from circular to elliptic paraboloids and then from incompressible to mildly compressible (subsonic) flow. The resulting spatial variations in density, which exhibit the expected pileup ahead of the stagnation point and a return to the undisturbed density at large lateral distances, are consistent with conservation of mass and approximately consistent with conservation of linear momentum along streamlines.
Even for this extension to compressibility, we found an exact solution of the associated magnetic field frozen into the flow at the expense of more involved, albeit still fully analytical expressions without recourse to special functions or unevaluated integrals.
Already in the simpler incompressible case, this property is unique among known analytical solutions for magnetized flow around solid obstacles, and therefore makes the present model a valuable tool not only in the magnetospheric context, but also for qualitative investigations of similar geometries such as the sphere/ellipsoid or the heliopause. \\

\section*{Acknowledgments}

\noindent JK gratefully acknowledges financial support from the German Research Foundation (\textit{Deutsche Forschungsgemeinschaft, DFG}) through Collaborative Research Center (\textit{Sonderforschungsbereich, SFB}) 1491.

\section*{Author declarations}

\noindent \textbf{Conflict of interest statement} \\
The authors have no conflicts to disclose. \\



\section*{Data availability}

\noindent Data sharing is not applicable to this article as no new data were created or analyzed in this study. \\

\appendix

\section{Range of admissible flow modulation factors}
\label{sec:appA}

\noindent We show that exact solutions of the irrotationality constraint (\ref{eq:irrotational}) for the particular ansatz~(\ref{eq:ansatz}) only exist if $\eps \in \{0, 1\}$. To this end, we insert the generalized stream function
\begin{equation}
  \label{eq:sfa}
  \Psi_f = - \frac{\rho^2}{2} \biggl(1 - \frac{1}{f(\rho, z)}\biggr)
\end{equation}
into the equation 
\begin{equation}
  \nabla \times \left[\nabla \times \left(\frac{\Psi_f}{\rho} \, \vec{e}_\phi \right) \right]= \vec{0} 
\end{equation}
already known from Section~\ref{sec:intro}, which yields 
\begin{equation}
  \partial_{\rho \rho} f - \frac{2}{f} \, (\partial_{\rho} f)^2 + \frac{3}{\rho} \, \partial_{\rho} f = - \partial_{z z} f + \frac{2}{f} \, (\partial_z f)^2 \, .
\end{equation}
Then, using Equation~(\ref{eq:ansatz}) and the new variable $\eta = r + \eps \, z$, we immediately obtain 
\begin{equation}
  \label{eq:irrcon} 
  \left(2 \, \eta + [\eps^2 - 1] \, r\right) \biggl[\partial_{\eta \eta} g_{\eps}(\eta) - \frac{2 \, [\partial_{\eta} g_{\eps}(\eta)]^2}{g_{\eps}(\eta)}\biggr] + 4 \, \partial_{\eta} g_{\eps}(\eta) = 0 \, .
\end{equation}
This nonlinear equation for $g_{\eps}$ can be reduced to a simpler linear equation for $h_{\eps} = 1/g_{\eps}$, namely
\begin{equation}
  \label{eq:simpirrcon} 
  \left(2 \, \eta + [\eps^2 - 1] \, r\right) \partial_{\eta \eta} h_{\eps}(\eta) + 4 \, \partial_{\eta} h_{\eps}(\eta) = 0 \, .
\end{equation}
Now, we find that for the parameter value $\eps = 0$, for which $\eta = r$, this equation further reduces to the ordinary differential equation 
\begin{equation}
  r \, \partial_{r r} h_0(r) + 4 \, \partial_r h_0(r) = 0 
\end{equation}
with the general solution
\begin{equation}
  h_0(r) = b + \frac{c}{r^3} 
\end{equation}
for $b, c \in \mathbb{R}$, reproducing Equation~(\ref{eq:psi_sphere}) if $b=0$ and $c=1$. The case $\eps = 1$, on the other hand, leads to an ordinary differential equation of the form
\begin{equation}
  \eta \, \partial_{\eta \eta} h_1(\eta) + 2 \, \partial_{\eta} h_1(\eta) = 0 
\end{equation}
with the general solution
\begin{equation}
  h_1(r + z) = b + \frac{c}{r + z} \, ,
\end{equation}
which gives rise to Equation~(\ref{eq:psi_para}) for the same choices of $b$ and $c$. Note that, since these constants of integration can be absorbed into the normalizations of length and velocity, different choices for $b$ and $c$ will not lead to qualitatively different solutions.

For the remaining parameter ranges $0 < \eps < 1$ and $\eps > 1$, it is obvious that Equation (\ref{eq:simpirrcon}) does not have any solutions, as the prefactor of the second-order contribution, in addition to $\eta$, now contains the independent variable $r$. Accordingly, general solutions $h_{\eps} \in C^2(\mathbb{R}, \mathbb{R})$ of the irrotationality constraint only exist for the cases $\eps = 0$ and $\eps = 1$. Employing a more general approach than the simple modulation factor ansatz (\ref{eq:sfa}) together with the conic section condition (\ref{eq:ansatz}) may, however, resolve this problem. \\

\section{Magnetic field components in other coordinate systems}
\label{sec:appB}
\setcounter{equation}{0}
 
\noindent To ease the application of our model, we summarize here, without derivation, the components of our magnetic field solution (for incompressible flow and homogeneous upstream boundary conditions) in cylindrical, Cartesian, and spherical coordinates. \\

\subsection{Cylindrical coordinates}

\noindent The cylindrical coordinate representation of our magnetic field solution is obtained by summing Equations~(\ref{eq:b_long}) and (\ref{eq:b_trans-1})--(\ref{eq:b_trans-3}). Written in terms of the Cartesian boundary components in Equation~(\ref{eq:Bxy_0}), it reads
\begin{align}
  \label{eq:final_cyl-1}
  B_\rho &= \sqrt{\frac{r+z}{r+z-1}} \left( 1 - \frac{r-z}{2 \, (r+z) \, r} \right) \left[ B_{x 0} \cos\phi + B_{y 0} \sin\phi \right]
  - \frac{1}{2 \, r} \sqrt{\frac{r-z}{r+z}} \, B_{z 0} \\[0.20cm]
  \label{eq:final_cyl-2}
  B_\phi &= \sqrt{\frac{r+z}{r+z-1}} \left[ -B_{x 0} \sin\phi + B_{y 0} \cos\phi \right] \\[0.20cm]
  \label{eq:final_cyl-3}
  B_z &= - \frac{1}{2 \, r} \, \sqrt{\frac{r-z}{r+z-1}} \left[ B_{x 0} \cos\phi + B_{y 0} \sin\phi \right] + \left(1-\frac{1}{2 \, r} \right) B_{z 0} \, .
\end{align}
For a parabolic obstacle shape that has been linearly scaled according to Equation~(\ref{eq:scaling_2d}), these components are generalized to
\begin{align}
 \label{eq:final_cyl_sc-1}
  B\sca_\rho &= \sqrt{\frac{r_\ell \, \ell_z +z}{(r_\ell-1) \, \ell_z + z}} \left( 1 - \frac{r_\ell \, \ell_z-z}{2 \, (r_\ell \, \ell_z+z) \, r_\ell} \right) \left[ B_{x 0} \cos\phi + B_{y 0} \sin\phi \right]
  - \frac{\ell_\rho}{2 \, r_\ell\, \ell_z} \sqrt{\frac{r_\ell \, \ell_z-z}{r_\ell \, \ell_z +z}} \, B_{z 0} \\[0.20cm]
  B\sca_\phi &= \sqrt{\frac{r_\ell \, \ell_z +z}{(r_\ell-1) \, \ell_z + z}} \left[ -B_{x 0} \sin\phi + B_{y 0} \cos\phi \right] \\[0.20cm]
  \label{eq:final_cyl_sc-3}
  B\sca_z &= - \frac{\ell_z}{2 \, r_\ell \, \ell_\rho} \, \sqrt{\frac{r_\ell \, \ell_z-z}{(r_\ell-1) \, \ell_z+z}} \left[ B_{x 0} \cos\phi + B_{y 0} \sin\phi \right] + \left(1-\frac{1}{2 \, r_\ell} \right) B_{z 0} \, ,
\end{align}
with $r_\ell = \sqrt{(\rho/\ell_\rho)^2+(z/\ell_z)^2}$. \\

\subsection{Cartesian coordinates}

\noindent The Cartesian coordinate representation can be found from the cylindrical components (\ref{eq:final_cyl-1})--(\ref{eq:final_cyl-3}) via the usual relations
\begin{equation}
  (B_x, B_y, B_z) = (B_\rho \cos\phi - B_\phi \sin\phi,  B_\rho \sin\phi + B_\phi \cos\phi, B_z) 
\end{equation}
and the substitutions $\cos\phi = x/\rho$, $\sin\phi = y/\rho$, and $\rho=\sqrt{x^2+y^2}=\sqrt{r^2-z^2}$ as
\begin{align}
  \label{eq:final_car-1}
 B_x &= \sqrt{\frac{r+z}{r+z-1}} \left( B_{x 0} - \frac{(x \, B_{x 0} + y \, B_{y 0}) \, x}{2 \, (r+z)^2 \, r} \right) - \frac{x \, B_{z 0}}{2 \, (r+z) \, r} \\[0.20cm]
  B_y &= \sqrt{\frac{r+z}{r+z-1}} \left( B_{y 0} - \frac{(x \, B_{x 0} + y \, B_{y 0}) \, y}{2 \,  (r+z)^2 \, r} \right) - \frac{y \, B_{z 0}}{2 \, (r+z) \, r} \\[0.20cm]
  \label{eq:final_car-3}
  B_z &= - \frac{x \, B_{x 0} + y \, B_{y 0}}{2 \, r \, \sqrt{(r+z)(r+z-1)}}
  + \left( 1-\frac{1}{2 \, r} \right) B_{z 0} 
\end{align}
after a series of elementary algebraic manipulations. Furthermore, when subjected to the triaxial scaling law of Equation~(\ref{eq:scaling_3d}), they become
\begin{align}
  \label{eq:final_car_sc-1}
  B_x &= \sqrt{\frac{r_\ell \, \ell_z+z}{(r_\ell-1)\, \ell_z +z}} \left( B_{x 0} - \frac{ \left[ \left(x/\ell_x^2\right) B_{x 0} + \left(y/\ell_y^2\right) B_{y 0} \right] x}{2 \, (r_\ell+z/\ell_z)^2 \, r_\ell} \right) - \frac{x \, B_{z 0}}{2 \, (r_\ell \, \ell_z+z) \, r_\ell} \\[0.20cm] 
  B_y &= \sqrt{\frac{r_\ell \, \ell_z+z}{(r_\ell-1)\, \ell_z +z}} \left( B_{y 0} - \frac{ \left[ \left(x/\ell_x^2\right) B_{x 0} + \left(y/\ell_y^2\right) B_{y 0} \right] y}{2 \, (r_\ell+z/\ell_z)^2 \, r_\ell} \right) - \frac{y \, B_{z 0}}{2 \, (r_\ell \, \ell_z+z) \, r_\ell} \\[0.20cm]
  \label{eq:final_car_sc-3}
  B_z &= - \frac{\left[ \left(x/\ell_x^2\right) B_{x 0} + \left(y/\ell_y^2\right) B_{y 0} \right] \ell_z^2 }{2 \, r_\ell \, \sqrt{\left(r_\ell \, \ell_z+z \right) \left( [r_\ell-1] \ell_z +z \right)}}
  + \left( 1-\frac{1}{2 \, r_\ell} \right) B_{z 0}
\end{align}
with $r_\ell$ now given by Equation~(\ref{eq:rL_car}).
Choosing $\ell_x = \ell_y = \ell_\rho$ reproduces the axisymmetrically scaled case~(\ref{eq:final_cyl_sc-1})--(\ref{eq:final_cyl_sc-3}), which in turn reduces to the standard case~(\ref{eq:final_cyl-1})--(\ref{eq:final_cyl-3}) of potential flow if $\ell_\rho = \ell_z$. \\

\subsection{Spherical coordinates}

\noindent For the sake of brevity, we content ourselves with the potential flow case of the spherical coordinate representation. This can be obtained from Equations~(\ref{eq:final_cyl-1})--(\ref{eq:final_cyl-3}) through the transformation
\begin{equation}
  (B_r, B_\tet, B_\phi) = (B_{\rho} \sin\tet + B_z \cos\tet, B_{\rho} \cos\tet - B_z \sin\tet, B_\phi)
\end{equation}
and the standard substitutions
$\rho = r \sin\tet$ and $z = r \cos\tet$, resulting in
\begin{align}
  B_r &= \frac{[2 \, (1+\cos\tet) \, r - 1] \, \sin\tet}{2 \, \sqrt{(1+\cos\tet) \, (r+r \cos\tet-1) \, r}} \, [B_{x 0} \cos\phi+B_{y 0 } \sin\phi]
  +\left( \cos\tet-\frac{1}{2 \, r} \right) B_{z 0} \\[0.20cm]
  B_\tet &= \frac{[ 2 \, (1+\cos\tet) \, r -1 ] \, \cos\tet +1}{2 \, \sqrt{(1+\cos\tet) \, (r+r\cos\tet-1) \, r}} \, [B_{x 0} \cos\phi+B_{y 0 } \sin\phi]
  - \sin\tet \left(1 -\frac{1}{2 \, (1+\cos\tet) \, r} \right) B_{z 0} \\[0.20cm]
  B_\phi &= \left(1- \frac{1}{(1+\cos\tet) \, r} \right)^{-1/2} [-B_{x 0} \sin\phi + B_{y 0} \cos\phi] \ .
\end{align}
~\\

\section{Determination of the functions $\vec{K}_1$ and $\vec{K}_2$}
\label{sec:appC}
\setcounter{equation}{0}

\noindent We present explicit analytical expressions for the functions $\vec{K}_1$ and $\vec{K}_2$ defined in Equation~(\ref{Ks}). Although we provide the formula for the function $\vec{K}_2$ for the sake of completeness, for many applications it will likely be sufficient to work with the first-order density approximation~(\ref{eq:nb_approx1}), whose magnetic field merely requires the simpler $\vec{K}_1$ but not the much more involved $\vec{K}_2$ (see below). In addition to its more complex formulas, a further potential complication with the second-order case may arise in numerical applications because several of the terms in $\vec{K}_2$ individually diverge in the vicinity of the $(a=0)$ streamline, and only produce values of order unity in their total sum through mutual cancellation. This is the main reason why the second-order magnetic field formulas have not been used in Figure~\ref{fig:fieldlines_2d}.

Suppressing the limit for the time being and using the Leibniz integral rule, we find that Equation (\ref{Ks}) may be written in the form
\begin{equation} \label{Kgen}
  \nabla \int_{\rho_0}^\rho \frac{\A^k}{u_\rho\inc} \, \dd\rho\p = \int_{\rho_0}^\rho \nabla \biggl(\frac{\A^k}{u_\rho\inc}\biggr) \, \dd\rho\p + \frac{\A^k}{u_\rho\inc} \, \nabla \rho - \frac{\A^k}{u_\rho\inc}\bigg|_{\rho' = \rho_0} \nabla \rho_0 \, ,
\end{equation}
where 
\begin{equation} \label{eq:rho0}
  \rho_0 = \rho_0(\rho, z; z_0) = \sqrt{\Bigl[\sqrt{(r - z) \, (r + z - 1) + (z_0 - 1/2)^2} + 1/2\Bigr]^2 - z_0^2} 
\end{equation}
is a direct result of the condition $a(\rho_0, z_0) = a(\rho, z)$. Evaluating the gradients on the right-hand side of (\ref{Kgen}), we obtain
\begin{equation} \label{IntK}
  \nabla \int_{\rho_0}^\rho \frac{\A^k}{u_\rho\inc} \, \dd\rho\p = \int_{\rho_0}^\rho \partial_{a} \biggl(\frac{\A^k}{u_\rho\inc}\biggr) \, \dd\rho\p \times \biggl[\frac{\partial a}{\partial \rho} \, \vec{e}_\rho + \frac{\partial a}{\partial z} \, \vec{e}_z\biggr] + \frac{\A^k}{u_\rho\inc} \, \vec{e}_\rho - \frac{\A^k}{u_\rho\inc}\bigg|_{\rho' = \rho_0} \, \biggl[\frac{\partial \rho_0}{\partial \rho} \, \vec{e}_\rho + \frac{\partial \rho_0}{\partial z} \, \vec{e}_z\biggr] \, .
\end{equation}
With 
\begin{equation}
  \frac{\A}{u_\rho\inc} = \rho + \frac{a^4}{(\rho^2 - a^2)^2 \, \rho} \, ,
\end{equation}
and therefore
\begin{equation}
  \partial_{a} \biggl(\frac{\A}{u_\rho\inc}\biggr) = \frac{4 \, a^3 \rho}{(\rho^2 - a^2)^3} \, , 
\end{equation}
the remaining integral on the right-hand side for the case $k = 1$ readily yields
\begin{equation}
  \int_{\rho_0}^\rho \partial_{a} \biggl(\frac{\A}{u_\rho\inc}\biggr) \, \dd\rho\p = \frac{(\rho^2 - \rho_0^2) \, (\rho^2 + \rho_0^2 -2 \, a^2) \, a^3}{(\rho^2 - a^2)^2 \, (\rho_0^2 - a^2)^2} \, . 
\end{equation}
Substituting this integral as well as Equation (\ref{eq:rel_a-Psi}) into Equation (\ref{IntK}) for $k = 1$ and performing the limit, $\vec{K}_1$ results in the expression 
\begin{equation} \label{eq:K1}
  \vec{K}_1 = \frac{1}{2 \, r} \, \sqrt{\frac{r - z}{r + z}} \,\, \vec{e}_\rho + \biggl(\frac{1}{2 \, r} - 1\biggr) \, \vec{e}_z \, .
\end{equation}

To obtain the expression for $\vec{K}_2$ that is required for the second-order approximation, we use
\begin{equation}
  \frac{\A^2}{u_\rho\inc} = \frac{\bigl[(\rho^2 - a^2)^2 \, \rho^2 + a^4\bigr]^2}{\bigl[\rho^2 + (\rho^2 - a^2)^2\bigr] \, (\rho^2 - a^2)^2 \, \rho^3} 
\end{equation}
and thus
\begin{equation}
  \partial_{a} \biggl(\frac{\A^2}{u_\rho\inc}\biggr) = \frac{4 \, \bigl[(\rho^2 - a^2)^2 \, \rho^2 + a^4\bigr] \, \bigl[(2 \, \rho^2 - a^2) \, a^2 - (\rho^2 - a^2)^2 \, (\rho^2 - 2 \, a^2)\bigr] \, a}{\bigl[\rho^2 + (\rho^2 - a^2)^2\bigr]^2 \, (\rho^2 - a^2)^3 \, \rho} \, ,
\end{equation}
such that the integral on the right-hand side of Equation~(\ref{IntK}) for $k = 2$ becomes
\begin{align} 
  \int_{\rho_0}^\rho \partial_{a} \biggl(\frac{\A^2}{u_\rho\inc}\biggr) \, \dd\rho\p =&\ - \frac{2}{a} \, \biggl(\frac{1}{a^4} - \frac{4}{a^2} + 2\biggr) \, Q - \frac{2 \, (\rho^2 - \rho_0^2) \, \bigl[\rho^2 \, \rho_0^2 + (1 - 3 \, a^2) \, (\rho^2 + \rho_0^2) - 5 \, (1 - a^2) \, a^2 + 1\bigr]}{\bigl[\rho^2 + (\rho^2 - a^2)^2\bigr] \, \bigl[\rho_0^2 + (\rho_0^2 - a^2)^2\bigr] \, a} \nonumber \\ \\
  & - \frac{(\rho^2 - \rho_0^2) \, \bigl[2 \, \rho^2 \, \rho_0^2 - 3 \, (\rho^2 + \rho_0^2) \, a^2 + 4 \, a^4\bigr] \, a}{(\rho^2 - a^2)^2 \, (\rho_0^2 - a^2)^2} + \frac{2 \, a^2 - 1}{a^5} \, \ln{\Biggl(\frac{\bigl[\rho^2 + (\rho^2 - a^2)^2\bigr] \, \rho_0^4}{\bigl[\rho_0^2 + (\rho_0^2 - a^2)^2\bigr] \, \rho^4}\Biggr)} \, , \nonumber 
\end{align}
where
\begin{equation} \label{fctQ}
  Q = Q(\rho, a; \rho_0) := \left\{ \begin{array}{lcl} \displaystyle
  \displaystyle \frac{1}{\sqrt{1 - 4 \, a^2}} \, \mathrm{arccoth}\Biggl(\frac{\rho^2 + \rho_0^2 + 2 \, (\rho^2 - a^2) \, (\rho_0^2 - a^2)}{\sqrt{1 - 4 \, a^2} \, (\rho^2 - \rho_0^2)}\Biggr) & \ : & \ a < 1/2   
  \\[0.65cm]
  \displaystyle \frac{8 \, (\rho^2 - \rho_0^2)}{(1 + 4 \, \rho^2) \, (1 + 4 \, \rho_0^2)} & \ : & \ a = 1/2 \\[0.50cm]
  \displaystyle \frac{1}{\sqrt{4 \, a^2 - 1}} \, \mathrm{arccot}\Biggl(\frac{\rho^2 + \rho_0^2 + 2 \, (\rho^2 - a^2) \, (\rho_0^2 - a^2)}{\sqrt{4 \, a^2 - 1} \, (\rho^2 - \rho_0^2)}\Biggr) & : \ & \ a > 1/2 \, .
  \end{array} \right.
\end{equation}
Again substituting the integral as well as Equations~(\ref{eq:rel_a-Psi}) and (\ref{eq:rho0}) into Equation~(\ref{IntK}) for $k = 2$ and performing the limit, the function $\vec{K}_2$ computes to
\begin{align} \label{eq:K2}
  \vec{K}_2 =& \frac{1}{2 \, r} \, \sqrt{\frac{r + z}{r - z}} \, \biggl[2 \, (r - 1) + \frac{1}{r + z}\biggr]^2 \, \vec{e}_\rho + \biggl[\biggl(1 - \frac{1}{2 \, r}\biggr) \, \sqrt{\frac{r + z}{r - z}} \ \vec{e}_{\rho} + \frac{1}{2 \, r} \, \vec{e}_z\biggr] \times \biggl[2 \, \biggl(1 - \frac{\rho^2}{a^2}\biggr) \, \biggl(\frac{1}{a^4} - \frac{4}{a^2} + 2\biggr) \biggr. Q(\rho, a; a) \nonumber \\[0.25cm]
  & \biggl. + 3 - 2 \, r - \frac{r^2 + 2 \, z^2 + 3 \, r \, z - 4 \, r - 5 \, z + 4 + 1/(r - z)}{(r + z - 1)^2 \, r} + \frac{(2 \, a^2 - 1) \, (r - z)}{a^6} \, \ln{\biggl(\frac{2 \, (r + z - 1) \, r}{(r + z)^2}\biggr)}\biggr] \, .
\end{align}
~\\

\section{The renormalized second Euler potential for the compressible case}
\label{sec:appD}
\setcounter{equation}{0}

\noindent We determine the renormalized second Euler potential
\begin{equation} \label{D1}
    \beta\com_\mathrm{ren} = \lim_{z_0\rightarrow\infty}
    \left[- \int_{\rho_0}^\rho \frac{\dd \rho\p}{u_\rho\com|_a}
    + \int_{z_0}^\za \frac{\dd z}{u_z\com|_{\rho=0}} \right]
    = \sum_{k=0}^2 \nu_k \lim_{z_0\rightarrow\infty} \big( \hat{\beta}_k + \Delta \beta_k \big) 
\end{equation}
for the case of compressible flow, in which
\begin{equation} \label{D2}
    \hat{\beta}_k := - \int_{\rho_0}^\rho
    \left. \frac{\A^k}{u_\rho\inc} \right|_a \dd \rho\p
    \quad  \mbox{and} \quad
    \Delta \beta_k := \int_{z_0}^\za
    \left. \frac{\A^k}{u_z\inc} \right|_{\rho=0} \dd z 
\end{equation}
with $\rho_0$ given by Equation~(\ref{eq:rho0}). The quantities $\hat{\beta}_0$ and $\Delta \beta_0$ defined in Equation (\ref{D2}) can be found in Sections~\ref{Sec-3-2} and \ref{sec:isochrones}. For $k \in \{1, 2\}$, the respective integrals yield

\begin{align}
 \hat{\beta}_1 =&\ - \int_{\rho_0}^\rho \left(
 \rho\p + \frac{a^4}{({\rho\p}^2-a^2)^2 \, \rho\p} \right) \dd\rho\p
 = \frac{1}{2} \Biggl[- {\rho\p}^2 + \frac{a^2}{{\rho\p}^2-a^2} - \ln \biggl( \frac{{\rho\p}^2}{{\rho\p}^2-a^2}\biggr) \Biggr]
  \Bigg|_{\rho_0}^\rho \\ \nonumber \\ \nonumber  
  \hat{\beta}_2 =&\ - \int_{\rho_0}^\rho \left(
  \frac{ \big[a^4 + ({\rho\p}^2-a^2)^2 \, {\rho\p}^2\big]^2}{{\big[{\rho\p}^2+({\rho\p}^2-a^2)^2 \big] \, ({\rho\p}^2-a^2)^2 \, \rho\p}^3} \right) \dd\rho\p = \biggl(\frac{1}{2 \, a^4} - \frac{3}{a^2} + 4\biggr) \, Q(\rho, a; \rho_0) \\[0.20cm]
  &\ + \frac{1}{2} \left[ \left. - {\rho\p}^2 + \frac{1}{{\rho\p}^2} + \frac{a^2}{{\rho\p}^2-a^2}
  + \ln \biggl( \frac{({\rho\p}^2-a^2)^3}{{\rho\p}^4} \biggr)
  - \frac{1}{a^2} \, \biggl(\frac{1}{2 \, a^2} - 2\biggr) \, \ln \biggl( 
  \frac{{\rho\p}^2 + ({\rho\p}^2-a^2)^2}{{\rho\p}^4} \biggr) \right] \right|_{\rho_0}^\rho \, ,
\end{align}
where the function $Q$ is specified in Equation (\ref{fctQ}), and
\begin{align}
   \Delta \beta_1 &= \int_{z_0}^\za \left( \frac{1}{2 \, z}-1 \right) \dd z
   = \left. \left[ \frac{\ln(z)}{2} - z \right] \right|_{z_0}^\za \\ \nonumber \\
   \Delta \beta_2 &= \int_{z_0}^\za \left( \frac{1}{2 \, z}-1 \right)^3 \dd z
   = \left. \left[ \frac{3 \, \ln(z)}{2} -z +\frac{3}{4\, z} -\frac{1}{16 \, z^2} \right] \right|_{z_0}^\za \, .
\end{align}
Then, the limits $L_k := \lim_{z_0\rightarrow\infty} \big( \hat{\beta}_k + \Delta \beta_k \big)$ read $L_0 = \beta\inc_\mathrm{ren}$ (cf.\ Equation (\ref{eq:beta_ren})) as well as
\begin{align}
    L_1 =&\ z - \frac{1}{2} \ln\left(\frac{r+z}{2}\right) \\ \nonumber \\
    L_2 =&\ z - \frac{1}{2} \, \ln \biggl(\frac{(r + z)^2 \, (r + z - 1)}{8} \biggr) - \biggl(\frac{1}{2 \, a^4} - \frac{3}{a^2} + 4\biggr) \, Q(\rho, a; a) - \frac{1}{2 \, (r + z - 1) \, \rho^2} \nonumber \\[0.20cm]
    &\ + \frac{(r - 1) \, r - (z - 1/2)^2}{(r - z)^2 \, (r + z - 1)^2} \, \ln \left( \frac{2 \, (r+z-1) \, r}{(r+z)^2}\right) + \frac{11}{16} \, ,  
\end{align}
with each $\za$ again adjusted such that $L_k|_{\rho=0,z=1}=1$ (cf.\ Section~ \ref{sec:isochrones}). Substituting these expressions together with $\nu_0 = \nsp$ into Equation (\ref{D1}), we are finally lead to
\begin{align}
    \label{eq:beta_ren_C}
   \beta\com_\mathrm{ren} =&\ \nonumber
    \underbrace{(\nsp+\nu_1+\nu_2)}_{=1} \, z + \frac{\nsp \ln (r + z - 1)}{2} - \frac{\nu_1}{2} \, \ln \left( \frac{r+z}{2} \right) + \nu_2 \, \Biggl[- \frac{1}{2} \, \ln \biggl(\frac{(r + z)^2 \, (r + z - 1)}{8} \biggr) \Biggr. \nonumber \\[0.20cm]
    &\ \Biggl. - \biggl(\frac{1}{2 \, a^4} - \frac{3}{a^2} + 4\biggr) \, Q(\rho, a; a) - \frac{1}{2 \, (r + z - 1) \, \rho^2} + \frac{(r - 1) \, r - (z - 1/2)^2}{(r - z)^2 \, (r + z - 1)^2} \, \ln \left( \frac{2 \, (r+z-1) \, r}{(r+z)^2}\right) + \frac{11}{16}\Biggr] \ .
\end{align}
Just as with the $\vec{K}_{1,2}$ terms in the magnetic field components, we see that the first-order case ($\nu_2 = 0$) is quite compact and devoid of apparent practical difficulties, whereas the second-order case ($\nu_2 \ne 0$) is not only more involved but contains several terms which individually tend to $\pm\infty$ as $a \rightarrow 0$ and only yield a finite result when considered together. This is most apparent in the final two terms in the square bracket of Equation~(\ref{eq:beta_ren_C}). \\

\bibliography{paraboloid}{}
\bibliographystyle{aasjournal}

\end{document}